\documentclass[prd,reprint,10pt,twoside,tightenlines,nofootinbib,showpacs,notitlepage,superscriptaddress,aps]{revtex4-1}
\pdfoutput=1
\usepackage{dcolumn}
\usepackage{amsmath,amsfonts,amssymb,latexsym}
\usepackage[sort&compress]{natbib}
\usepackage{grffile,epstopdf}
\usepackage{graphicx}
\usepackage{hhline}
\usepackage{xcolor}
\usepackage{hhline}
\usepackage{physics}

\makeindex

\begin{document}
	
\title{Diffusion coefficient matrix of the strongly interacting quark-gluon plasma}
	
\author{Jan A. Fotakis}
	\email{fotakis@itp.uni-frankfurt.de}
	\affiliation{Institut f\"ur Theoretische Physik, Johann Wolfgang Goethe-Universit\"at,
		Max-von-Laue-Str.\ 1, D-60438 Frankfurt am Main, Germany}
	
\author{Olga Soloveva}
%	\email{soloveva@itp.uni-frankfurt.de}
\affiliation{Institut f\"ur Theoretische Physik, Johann Wolfgang Goethe-Universit\"at,
		Max-von-Laue-Str.\ 1, D-60438 Frankfurt am Main, Germany}

\author{Carsten Greiner}
	\affiliation{Institut f\"ur Theoretische Physik, Johann Wolfgang Goethe-Universit\"at,
		Max-von-Laue-Str.\ 1, D-60438 Frankfurt am Main, Germany}
		
\author{Olaf Kaczmarek}
	\affiliation{Fakult\"at f\"ur Physik, Universit\"at Bielefeld, D-33615  Bielefeld, Germany}
	\affiliation{Key Laboratory of Quark and Lepton Physics (MOE) and Institute of Particle Physics, Central China Normal University, Wuhan 430079, China}	
\author{Elena Bratkovskaya}
	\affiliation{GSI Helmholtzzentrum f\"ur Schwerionenforschung GmbH,
		Planckstrasse 1, D-64291 Darmstadt, Germany}
	\affiliation{Institut f\"ur Theoretische Physik, Johann Wolfgang Goethe-Universit\"at,
		Max-von-Laue-Str.\ 1, D-60438 Frankfurt am Main, Germany}
	
	\date{\today }
	
	%%%%%%%%%%%%%%%%%%%%%%%%%%%%%%%%%%%%%%%%%%%%%%%%%%%%%%%%%%%%%%
	%%%%%%%%%%%%%%%%%%%%%%%%%%%%%%%%%%%%%%%%%%%%%%%%%%%%%%%%%%%%%%
	% A B S T R A C T
	%%%%%%%%%%%%%%%%%%%%%%%%%%%%%%%%%%%%%%%%%%%%%%%%%%%%%%%%%%%%%%
	%%%%%%%%%%%%%%%%%%%%%%%%%%%%%%%%%%%%%%%%%%%%%%%%%%%%%%%%%%%%%%
	\begin{abstract}
We study the diffusion properties of the strongly interacting quark-gluon plasma (sQGP)
and evaluate the diffusion coefficient matrix  for the baryon $(\mathrm{B})$, strange $(\mathrm{S})$ and electric $(\mathrm{Q})$ charges - $\kappa_{qq^\prime}$  ($q,q^\prime=\mathrm{B}, \mathrm{S}, \mathrm{Q}$) and show their dependence
on temperature $T$ and baryon chemical potential $\mu_\mathrm{B}$.
The non-perturbative nature of the sQGP is evaluated within the Dynamical Quasi-Particle Model (DQPM) which is matched to reproduce the 
equation of state of the partonic matter  above the deconfinement temperature $T_c$ 
from lattice QCD. The calculation of diffusion coefficients is based on two methods:
i) the Chapman-Enskog method for the linearized Boltzmann equation, 
which allows to explore non-equilibrium corrections for the phase-space distribution function in leading order of the Knudsen numbers as well as ii) 
the relaxation time approximation (RTA). In this work we explore the differences between 
the two methods. We find a good agreement with the available lattice QCD data in case of the electric charge diffusion coefficient (or electric conductivity) at vanishing baryon chemical potential as well as a qualitative agreement with the recent predictions from the holographic approach for all diagonal components of the diffusion coefficient matrix.
The knowledge of the diffusion coefficient matrix is also of special interest for  
more accurate hydrodynamic simulations.
	\end{abstract}
	
	\maketitle

	%%%%%%%%%%%%%%%%%%%%%%%%%%%%%%%%%%%%%%%%%%%%%%%%%%%%%%%%%%%%%%
	%%%%%%%%%%%%%%%%%%%%%%%%%%%%%%%%%%%%%%%%%%%%%%%%%%%%%%%%%%%%%%
	% I N T R O D U C T I O N
	%%%%%%%%%%%%%%%%%%%%%%%%%%%%%%%%%%%%%%%%%%%%%%%%%%%%%%%%%%%%%%
	%%%%%%%%%%%%%%%%%%%%%%%%%%%%%%%%%%%%%%%%%%%%%%%%%%%%%%%%%%%%%%
	\section{Introduction}
	\label{sec:Intro}

An exploration of the properties of hot and dense matter - created in heavy-ion
collisions (HICs) at relativistic energies - is in the focus of extensive  research.  
It is the primary goal of experimental programs of the LHC (Large-Hadron-Collider) 
at CERN, the RHIC (Relativistic Heavy Ion Collider) at BNL, 
the future FAIR (Facility for Antiproton and Ion Research) at GSI, and  the NICA 
(Nuclotron-based Ion Collider fAcility) facility at JINR, which reproduce in the
laboratory the extreme conditions of the early stages of our universe by 'tiny bangs'. 
In the central region of heavy-ion collisions the deconfined QCD (Quantum  Chromo Dynamics)  matter -- a quark-gluon plasma (QGP) - is created 
which can achieve an approximate local equilibrium and exhibit 
hydrodynamic flow \cite{Ollitrault:1992bk,Heinz:2001xi,Shuryak:2008eq}.
The hydrodynamic behaviour of the fluid can be characterized by transport coefficients
such  as shear $\eta$, bulk $\zeta$ viscosities, and diffusion coefficients 
$\kappa$, which describe the fluid’s dissipative corrections at leading order. 
The interpretation of the experimental data, and especially the elliptic flow $v_2$, 
in terms of the hydrodynamic models showed that the QGP behaves almost as a nearly perfect fluid with a very low shear viscosity to entropy density ($s$) ratio, 
$\eta/s$, which reflects that its properties correspond to nonperturbative, 
strongly interacting matter \cite{Shuryak:2004cy,Gyulassy:2004zy,Romatschke:2007mq}.

By performing an experimental energy scan of HICs one can explore the different stages 
of the QCD phase diagram. At ultra-relativistic heavy-ion collisions 
at LHC and RHIC energies, the QGP is created at very large temperatures $T$ 
and almost zero or low baryon chemical potential $\mu_\mathrm{B}$, where according to 
lattice QCD (lQCD) results \cite{Cheng:2007jq,Aoki:2009sc} the transition from 
the QGP to the hadronic matter is a crossover.
By reducing the collision energy one can also explore the large $\mu_\mathrm{B}$ region
where one might expect the existence of a critical point and a 1st order phase
transition. Such conditions are presently under investigation within the  
RHIC BES (Beam Energy Scan) experiments and in future by the FAIR and NICA facilities.

The theoretical description of the QCD matter at finite $\mu_\mathrm{B}$, 
and especially in the vicinity of the critical point, requires an appropriate
description of transport of conserved charges -- baryon $B$, strangeness $S$
and electric $Q$ charges.   
In order to study  
the phenomenon of baryon-stopping, the baryon diffusion was recently introduced to various fluid dynamic models \cite{Denicol2018a,Li2018,Du:2019obx}. 
Moreover, the baryon diffusion coefficient has been studied in Refs. 
\cite{Arnold:2003zc,Rougemont:2015ona,Greif2017PRL,Ghiglieri:2018dib,Fotakis:2019nbq,Soloveva:2019xph}. 

In the recent past we have addressed the coupling of the conserved baryon number, strangeness and electric charge; the diffusion coefficient matrix 
($\kappa_{qq^\prime}$, where $q,q^\prime=\mathrm{B}, \mathrm{S}, \mathrm{Q}$)
was introduced and evaluated for a hadron gas and a simple model for quark-gluon plasma (QGP) \cite{Greif2017PRL,Fotakis:2019nbq}. These investigations were followed by a more extended study in the hadronic phase from kinetic theory in the case of the electric cross-conductivities \cite{Rose:2020sjv}. Furthermore, a first study on the impact of the coupling of baryon number and strangeness was provided in Ref. \cite{Fotakis:2019nbq}.
For this the diffusion coefficient matrix of hot and dense nuclear matter has to be investigated thoroughly being important for accurate hydrodynamic simulations. It was further motivated that the off-diagonal coefficients may have implications on the chemical-composition of the hadronic phase \cite{Rose:2020sjv}. 

The study of the transport of conserved electric charge $Q$ during heavy-ion collisions has been in the focus of intensive research. Due to its importance for the description of soft photon spectra and rates \cite{Turbide:2004PRC,Akamatsu:2011,PHSDreview,Yin:2014PRC} as well as for hydrodynamic approaches modelling the generation and  evolution of electromagnetic fields \cite{Tuchin:2013,Inghirami:2019mkc,Denicol:2019iyh,Oliva:2020mfr}, much attention was paid to the electric conductivity within  different theoretical approaches for the evaluation of the properties of the partonic and hadronic matter \cite{Arnold:2000dr,Brandt:2012jc,Torres-Rincon:2012sda,Finazzo:2013efa,Amato:2013naa,RMarty:2013xph,Aarts:2014nba,Greif:2014oia,Puglisi:2014sha,PUGLISI:2015PLB,Brandt:2015aqk,Rougemont:2015ona,Ding:2016hua,Greif:2016skc,L.Thakur:17PRD,Rougemont:2017tlu,Greif2017PRL,Hammelmann:2018ath,Soloveva:2019xph,Fotakis:2019nbq,Rose:2020sjv}. 
	
The exploration of the QGP properties at finite ($T,\mu_\mathrm{B}$) 
are of special interest for an understanding of the phase transition. 
The transport properties of the strongly interacting QGP
has been studied using the Dynamical Quasi-Particle Model(DQPM) 
\cite{Peshier:2005pp,Cassing:2007nb,Cassing:2007yg,Linnyk:2015rco,Berrehrah:2016vzw}
that is matched to reproduce the equation of state of the partonic system 
above the deconfinement temperature $T_c$ from lattice QCD. 
The DQPM is based on a propagator representation with complex self energies which
describes the degrees of freedom of the QGP in terms of strongly interacting  
dynamical quasiparticles which reflect  the non-perturbative nature of the QCD 
in the vicinity of the phase transition
where the QCD coupling grows rapidly with decreasing temperature according to 
lQCD calculations \cite{Kaczmarek:2004gv}.
Moreover, the DQPM allows to explore the properties of the QGP at finite $(T,\mu_\mathrm{B})$, expressed 
in terms of transport coefficients such as shear $\eta$, bulk $\zeta$ viscosities, 
baryon diffusion coefficients $\kappa_B$ and electric conductivity $\sigma_0$ 
based on the RTA (relaxation time approximation) \cite{Berrehrah:2016vzw,Moreau:2019vhw,Soloveva:2019xph}.

We note that an important advantage of a propagator based approach is that one 
can formulate a consistent thermodynamics~\cite{Vanderheyden:1998ph} and
a causal theory for non-equilibrium dynamics on the basis of Kadanoff--Baym 
equations \cite{KadanoffBaym}.
This allows to use the DQPM for the description of the partonic interactions 
and parton properties in the microscopic  Parton--Hadron--String Dynamics (PHSD)
transport approach
\cite{Cassing:2008sv,Cassing:2008nn,Cassing:2009vt,Bratkovskaya:2011wp,Linnyk:2015rco}
and to study the QGP properties out-of equilibrium as created in HICs as well as
in equilibrium by performing box calculations with periodic boundary conditions  \cite{Ozvenchuk:2012fn}. Moreover, the $(T,\mu_\mathrm{B})$ dependence of partonic 
properties and interaction cross sections have been explored in a more recent study within PHSD 5.0 \cite{Moreau:2019vhw,Soloveva:2019xph,Soloveva:2020ozg,Moreau:2021clr}.

We note that the studies of transport coefficients ($\eta/s, \zeta/s, \kappa_B, \sigma_0$)
within the DQPM (and PHSD) has been based on the relaxation-time approximation (RTA)
as incorporated in Refs. \cite{Hosoya:1983xm,Chakraborty:2010fr,Albright:2015fpa,Gavin:1985ph}
as well as  on the Kubo formalism~\cite{Kubo:1957mj,Aarts:2002cc,FernandezFraile:2005ka,Lang:2015nca} for $\eta/s$ (cf. \cite{Ozvenchuk:2012fn,Moreau:2019vhw}).
In Refs. \cite{Greif:2016skc,Greif2017PRL,Fotakis:2019nbq} the evaluation of the 
diffusion coefficient matrix has been done within  the Chapman-Enskog method  \cite{chapman1970mathematical} which allows to explore non-equilibrium corrections for the phase-space distribution function in leading order of the Knudsen numbers.

In the present study we combine the developments of Refs. 
\cite{Greif2017PRL,Fotakis:2019nbq,Soloveva:2019xph} and  evaluate  
the diffusion coefficient matrix of the strongly interacting non-perturbative QGP
at finite ($T,\mu_\mathrm{B}$), with properties described by the DQPM model, based on
recently explored the Chapman-Enskog method \cite{Greif:2016skc,Greif2017PRL,Fotakis:2019nbq}. This allows us to explore the influence of traces of 
non-equilibrium effects by accounting for the higher modes of the distribution function 
on the transport properties and compare the results with the often used kinetic RTA approximation. 
We provide the ($T,\mu_\mathrm{B}$) dependence of the diffusion coefficients $\kappa_{qq^\prime}$
for $q,q^\prime=\mathrm{B}, \mathrm{S}, \mathrm{Q}$ charges for  baryon chemical potentials $\mu_\mathrm{B} \leq 0.5$ GeV, where the phase transition is a rapid crossover.

This paper is structured as follows. In Section \ref{sec:Found} we provide 
a short review of the basic definitions and conventions, followed by a 
reminder of the basic ideas of the Chapman-Enskog method and its relaxation time approximation, which was used to evaluate the diffusion coefficient matrix in Refs. \cite{Greif:2016skc,Greif2017PRL,Fotakis:2019nbq}, and a short review of the dynamical quasi-particle model (DQPM) \cite{Peshier:2005pp,Cassing:2007nb,Cassing:2007yg,Cassing:2008nn,Soloveva:2019xph} in Section \ref{subsec:DQPM}. In the preface of Section \ref{sec:Results} we explain how to achieve results for the diffusion matrix from the DQPM by using the Chapman-Enskog method, and we demonstrate the differences between various assumptions in Section \ref{subsec:ConstCross} by providing an simple example. Finally, we provide and discuss improved results for all diffusion coefficients and conductivities and compare them to the available results from other approaches.

	%%%%%%%%%%%%%%%%%%%%%%%%%%%%%%%%%%%%%%%%%%%%%%%%%%%%%%%%%%%%%%
	%%%%%%%%%%%%%%%%%%%%%%%%%%%%%%%%%%%%%%%%%%%%%%%%%%%%%%%%%%%%%%
	% B A S I C  D E F I N I T I O N S
	%%%%%%%%%%%%%%%%%%%%%%%%%%%%%%%%%%%%%%%%%%%%%%%%%%%%%%%%%%%%%%
	%%%%%%%%%%%%%%%%%%%%%%%%%%%%%%%%%%%%%%%%%%%%%%%%%%%%%%%%%%%%%%
%----------------------------------------------------------------------	
	\section{Foundations}
	\label{sec:Found}
	
	Let $x \equiv x^\mu$ be the four-coordinate and $k \equiv k^\mu$ the four-momentum. The single-particle distribution function, $f_{i,\mathbf{k}} \equiv f_{i}(x,k)$, of a multi-component quasi-particle system obeys the effective Boltzmann equation \cite{Romatschke:2011qp}
	\begin{align}
	k_i^{\mu }\partial_\mu f_{i,\mathbf{k}} + \frac{1}{2}\partial_\mu \left( m_i^2 \right) \frac{\partial}{\partial k_{i,\, \mu}} f_{i,\mathbf{k}} = \sum\limits_{j \, = \, 1}^{N_{\text{species}}}C_{ij}(x,k), \label{eq:BoltzmannEq}
	\end{align}
	where $C_{ij}$ is the collision term and the masses depend on temperature and  chemical potentials, i.e. $m_i \equiv m_i(T, \mu_{\mathrm{B}}, \mu_{\mathrm{Q}}, \mu_{\mathrm{S}})$. The (local) equilibrium state of the system is described by
	\begin{align}
	f_{i,\mathbf{k}}^{(0)} = \frac{ g_{i} }{\exp \left( u_{\mu }k_i^{\mu }/T - \mu_i/T\right) - a_i }  , \label{eq:Equilibrium}
	\end{align}
	where $\mu_{i}=B_{i}\mu_\mathrm{B} + Q_{i}\mu_\mathrm{Q} + S_{i}\mu_\mathrm{S}$ is the chemical
	potential, $g_i$ is the degeneracy of the $i$-th species and 
	\begin{align}
	a_i \equiv \begin{cases}
	+1 & \text{(Bosons)}, \\ -1 & \text{(Fermions)}, \\ 0 & \text{(Classical)}.
	\end{cases}
	\end{align}
	Further, we define in short hand notation:
	\begin{align}
		\tilde{f}_{i,\mathbf{k}}^{(0)} \equiv 1 + a_i \frac{f_{i,\mathbf{k}}^{(0)}}{g_i} = 1 +  \frac{ a_i }{\exp \left( u_{\mu }k_i^{\mu }/T - \mu_i/T\right) - a_i } .
	\label{eq:Equil_tilde_f0}
	\end{align}
	Furthermore, the isotropic local equilibrium pressure is determined by the temperature and chemical potentials, $P_0 \equiv P_0\left(T, \mu_{\mathrm{B}}, \mu_{\mathrm{Q}}, \mu_{\mathrm{S}} \right)$. In this work, we adapt the isotropic pressure from lattice QCD \cite{Borsanyi:2012cr,Borsanyi:2013bia}. From the equation of state the energy density and the net charge densities are defined:
	\begin{align}
		\epsilon \equiv \epsilon\left(T, \mu_{\mathrm{B}}, \mu_{\mathrm{Q}}, \mu_{\mathrm{S}} \right), \quad n_q \equiv n_q\left(T, \mu_{\mathrm{B}}, \mu_{\mathrm{Q}}, \mu_{\mathrm{S}} \right),\nonumber \\
		 q \in \lbrace \mathrm{B},\mathrm{Q},\mathrm{S} \rbrace .
	\end{align}
	In kinetic theory the net charge densities are defined as:
	\begin{align}
		n_q = \sum\limits_{i \, = \, 1}^{N_{\text{species}}} q_i \int \mathrm{d} K_i \, E_{i,\mathbf{k}} f^{(0)}_{i,\mathbf{k}}, \quad q \in \lbrace \mathrm{B},\mathrm{Q},\mathrm{S} \rbrace \label{eq:chargedensities}
	\end{align}
	where $q$ is the type of the conserved quantum number, i.e. namely baryon number $\mathrm{B}$, strangeness $\mathrm{S}$ or electric charge $\mathrm{Q}$, and $q_i$ is the quantum number (of type $q$) of the $i$-th species. In this work we assume a partonic system with three flavors and thus the following particle species: up- ($u$), down- ($d$), strange-quark ($s$), the gluon ($g$), and the corresponding anti-particles. 
Furthermore, the Landau matching conditions were assumed \cite{landau1959course}:
	\begin{align}
		\sum\limits_{i \, = \, 1}^{N_{\text{species}}} q_i \int \mathrm{d} K_i \, E_{i,\mathbf{k}} \left( f_{i,\mathbf{k}} - f^{(0)}_{i,\mathbf{k}} \right) = 0,
	\end{align}
using the notation
	\begin{align}
		\mathrm{d}K_i \equiv \frac{\mathrm{d}^3 \mathbf{k}_i}{(2\pi)^3 E_{i,\mathbf{k}}},
	\end{align}
with the on-shell energy $E_{i,\mathbf{k}} = \sqrt{m_i^2 + \mathbf{k}_i^2}$.
	
	An (unpolarized) interaction is characterized by the invariant matrix-element $\bar{\mathcal{M}}_{i_1 \dots i_n \rightarrow j_1 \dots j_m} \equiv \bar{\mathcal{M}}(k_{i_1} \dots k_{i_n} \rightarrow p_{j_1} \dots p_{j_m})$, which is averaged over the ingoing spin-states and is summed over the outgoing spin-states. The differential cross section for a binary process of on-shell particles ($ i + j \rightarrow a + b$) in the center-of-momentum frame (CM), where the momenta of the colliding particles obey $\mathbf{k}_i + \mathbf{k^\prime}_j = \mathbf{p}_a + \mathbf{p^\prime}_b = \mathbf{P} = 0$ and $k^0_i + {k^\prime}^0_j = \sqrt{s} = p^0_i + {p^\prime}^0_j$, is given by	
	\begin{equation}
	\mathrm{d}\sigma_{ij \rightarrow ab}(\sqrt{s},\Omega) = \frac{1}{64 \pi^2 s} \frac{p_{\mathrm{out}}}{p_\mathrm{in}} |\bar{\mathcal{M}}|^2 \mathrm{d}\Omega,
	\label{dsigma_on_CM}
	\end{equation}
	where $s$ in the Mandelstam variable and $\mathrm{d}\Omega$ is the differential solid angle corresponding to one of the final particles. The momenta of the initial ($p_\mathrm{in}$) and final particles ($p_\mathrm{out}$) in the CM frame are found to be:
	\begin{equation}
	p_{i} = \frac{\sqrt{\left(s-(m_{i} + m^\prime_{i})^2\right)\left(s-(m_{i}-m^\prime_{i})^2\right)}}{2\sqrt{s}} ,
	\end{equation}
	~\\
	where $i= \mathrm{in}/\mathrm{out}$, $m_{i}$ and $m^\prime_{i}$ being the masses of the colliding partons. The total cross section is obtained via:
	\begin{align}
	\sigma^{ij \rightarrow ab}_{\mathrm{tot}}(\sqrt{s}) &\equiv 2\pi \gamma_{ij} \int \mathrm{d}\cos(\vartheta) \, \frac{\mathrm{d}}{\mathrm{d}\Omega} \sigma_{ij \rightarrow ab}(\sqrt{s},\cos(\vartheta)) \nonumber \\ 
	&= \frac{1}{32 \pi s} \frac{p_\mathrm{out}}{p_\mathrm{in}} \gamma_{ij} \int_{-1}^{1} \mathrm{d} \cos(\vartheta) \,  |\bar{\mathcal{M}}|^2 ,
	\label{sigma_on_CM}
	\end{align}
	~\\
	where $\vartheta$ is the final polar angle of one of the final particles in the CM frame, and $\gamma_{ij} = 1 - \frac{1}{2}\delta_{ij}$ is the symmetry factor.
	
	In this paper we use the short-hand $``\lbrace \mu_q \rbrace"$ instead of $``\mu_{\mathrm{B}}, \mu_{\mathrm{Q}}, \mu_{\mathrm{S}}"$ in function arguments, and the $(+,-,-,-)$-signature for the metric. Greek indices run from 0 to 3 and latin ones run from 1 to 3. Furthermore, we employ natural units, $\hbar = c = k_{\mathrm{B}} = 1$.
	
%----------------------------------------------------------------------	
	\subsection{First-order Chapman-Enskog approximation}
	\label{subsec:CE}
	
	If the perturbations from equilibrium are small, one can expand the single-particle distribution function in orders of the Knudsen number ($\mathrm{Kn}$):
	\begin{align}
		f_{i,\mathbf{k}} = f^{(0)}_{i,\mathbf{k}} + \epsilon f^{(1)}_{i,\mathbf{k}} + \mathcal{O}(\epsilon^2),
	\end{align}
	where $\epsilon$ is an assisting parameter for counting the orders of the gradients (or equivalently, the orders
	 of the Knudsen number), which will be send to 1 afterwards. This approximation is known as the Chapman-Enskog expansion to first order (CE) \cite{chapman1970mathematical}. Neglecting second-order terms leads to the linearized effective Boltzmann equation:
	 \begin{align}
	 	k_i^{\mu }\partial_\mu f^{(0)}_{i,\mathbf{k}} + \frac{1}{2}\partial_\mu \left( m_i^2 \right) \frac{\partial}{\partial k_{i,\, \mu}} f^{(0)}_{i,\mathbf{k}} = \sum\limits_{j \, = \, 1}^{N_{\text{species}}} \mathcal{C}_{ij}^{(1)}[f_{i,\textbf{k}}], \label{eq:LinEffBoltzmannEq}
	 \end{align}
	with the linearized collision term
%\begin{widetext}
\begin{align}
& \sum\limits_{j \, = \, 1}^{N_{\text{species}}} \mathcal{C}_{ij}^{(1)}[f_{i,\textbf{k}}] \equiv \frac{1}{2} \sum\limits_{j,a,b \, = \, 1}^{N_{\text{species}}} \int_{\mathbb{R}^3} \mathrm{d}P_a \int_{\mathbb{R}^3} \mathrm{d}P^\prime_b \int_{\mathbb{R}^3} \mathrm{d}K^\prime_j \, \nonumber \\
& \times W^{ij \rightarrow ab}_{k k^\prime \rightarrow p p^\prime} f^{(0)}_{i,\textbf{k}} f^{(0)}_{j,\textbf{k}^\prime} \tilde{f}^{(0)}_{a,\textbf{p}} \tilde{f}^{(0)}_{b,\textbf{p}^\prime}  \nonumber \\
 & \times  \left( \frac{f^{(1)}_{a,\textbf{p}}}{f^{(0)}_{a,\textbf{p}} \tilde{f}^{(0)}_{a,\textbf{p}}} + \frac{f^{(1)}_{b,\textbf{p}^\prime}}{f^{(0)}_{b,\textbf{p}^\prime} \tilde{f}^{(0)}_{b,\textbf{p}^\prime}}  - \frac{f^{(1)}_{i,\textbf{k}}}{f^{(0)}_{i,\textbf{k}} \tilde{f}^{(0)}_{i,\textbf{k}}} - \frac{f^{(1)}_{j,\textbf{k}^\prime}}{f^{(0)}_{j,\textbf{k}^\prime} \tilde{f}^{(0)}_{j,\textbf{k}^\prime}} \right), \label{eq:LinCollTermGeneral}
\end{align}
%\end{widetext}
	and 
	\begin{align}
		W^{ij \rightarrow ab}_{k k^\prime \rightarrow p p^\prime} \equiv \frac{(2\pi)^4}{16} \delta^{(4)}\left( k_i + k^\prime_j - p_a - p^\prime_b \right) \abs{ \bar{\mathcal{M}}_{ij \rightarrow ab} }^2
	\end{align}
for the inelastic binary transition rate. To linear order the diffusion currents are given via:
	\begin{align}
		V^\mu_q =  \sum\limits_{i \, = \, 1}^{N_{\text{species}}} q_i \int_{\mathbb{R}^3} \mathrm{d}K_i \, k^{\langle \mu \rangle}_i f^{(1)}_{i,\mathbf{k}},
	\end{align}
	and therefore the explicit mass-term in Eq. \eqref{eq:LinEffBoltzmannEq} does not affect the currents due to the anti-symmetry of the integrand \cite{Fotakis:2019nbq}. Here $q_i$ is again the quantum number of type $q \in \lbrace \mathrm{B},\mathrm{S},\mathrm{Q} \rbrace$ of the $i$-th particle species. 
	
For further evaluations with the CE method in this study we consider a classical system of on-shell particles, $a_i = 0 ~ \forall i$, and elastic binary collisions only, such that the on-shell transition rate for this case reads:
	\begin{align}
		&W^{ij \rightarrow ab}_{k k^\prime \rightarrow p p^\prime} = \gamma_{ij} (\delta_{ia}\delta_{jb} + \delta_{ib}\delta_{ja}) (2\pi)^6 s\, \nonumber \\
		&\times \left( \frac{\mathrm{d}}{\mathrm{d}\Omega}\sigma_{ij \rightarrow ab}(\sqrt{s},\Omega) \right) \delta^{(4)}\left(k_i + k^\prime_j - p_a - p^\prime_b\right) . \label{eq:ElasticTransitionRate}
	\end{align}
Additionally, we assume isotropic scattering processes. 
(The underlying assumptions will be discussed also in Section III.)
	
	Later we will incorporate  total cross sections for elastic binary processes originating from the dynamic quasi-particle model (DQPM) \cite{Soloveva:2019xph} 
which depend on temperature and baryon-chemical potential, $\sigma_{\mathrm{tot}}^{ij \rightarrow ab} \equiv \sigma_{\mathrm{tot}}^{ij \rightarrow ab}(\sqrt{s}, T, \mu_{\mathrm{B}})$ (Section \ref{subsec:DQPM}). \newline 
	Following the steps taken in Refs. \cite{Greif:2016skc,Greif2017PRL,Fotakis:2019nbq}, we can express the diffusion coefficient matrix for a classical system under the assumption of elastic isotropic scattering processes as
	\begin{align}
		\kappa_{qq^\prime} = \frac{1}{3} \sum\limits_{i \, = \, 1}^{N_{\text{species}}} q_i \sum\limits_{m \, = \, 0}^M \lambda^{(i)}_{m,q^\prime} \int_{\mathbb{R}^3} \mathrm{d}K_i \, E^m_{i,\mathbf{k}} \left( m_i^2 - E^2_{i,\mathbf{k}} \right) f^{(0)}_{i,\mathbf{k}}, 
		\label{eq:DiffMatrLin}
	\end{align}
	where the scalars $\lambda^{(i)}_{m,q^\prime}$ are solutions of the linearized Boltzmann equation in the form \cite{Greif:2016skc,Greif2017PRL,Fotakis:2019nbq}:
	\begin{align}
	\sum\limits_{m \, = \, 0}^M \sum\limits_{j \, = \, 1}^{N_{\text{species}}} \left( \mathcal{A}^{i}_{nm} \delta^{ij} + \mathcal{C}^{ij}_{nm} \right) \lambda^{(j)}_{m,q} = b^i_{q,n}, \label{eq:FinalSetLinEq}
	\end{align}
	with the abbreviations
	\begin{widetext}
		\begin{align}
			\mathcal{A}^i_{nm} &\equiv \sum\limits_{\ell \, = \, 1}^{N_{\text{species}}} \gamma_{i\ell} \int \mathrm{d}K_i \mathrm{d}K^\prime_\ell \mathrm{d}P_i \mathrm{d}P^\prime_\ell \, (2\pi)^6 s \left( \frac{\mathrm{d}}{\mathrm{d}\Omega}\sigma_{i\ell \rightarrow i\ell} \right) \, \delta^{(4)}\left(k_i + k^\prime_\ell - p_i - p^\prime_\ell\right) f^{(0)}_{i,\textbf{k}} f^{(0)}_{\ell,\textbf{k}^\prime} E_{i,\textbf{k}}^{n-1} k_{i,\,\langle \alpha \rangle} \left( E^m_{i,\textbf{p}} p^{\langle \alpha \rangle}_i - E^m_{i,\textbf{k}} k^{\langle \alpha \rangle}_i \right), \nonumber \\
			\mathcal{C}^{ij}_{nm} &\equiv \gamma_{ij} \int \mathrm{d}K_i \mathrm{d}K^\prime_j \mathrm{d}P_i \mathrm{d}P^\prime_j \, (2\pi)^6 s \left( \frac{\mathrm{d}}{\mathrm{d}\Omega}\sigma_{ij \rightarrow ij} \right) \, \delta^{(4)}\left(k_i + k^\prime_j - p_i - p^\prime_j\right) f^{(0)}_{i,\textbf{k}} f^{(0)}_{j,\textbf{k}^\prime} E_{i,\textbf{k}}^{n-1} k_{i,\,\langle \alpha \rangle} \left( E^m_{j,\textbf{p}^\prime} {p^\prime}^{\langle \alpha \rangle}_j - E^m_{j,\textbf{k}^\prime} {k^\prime}^{\langle \alpha \rangle}_j \right), \nonumber\\
		b^i_{q,n} &\equiv \int_{\mathbb{R}^3} \mathrm{d}K_i \, E_{i,\textbf{k}}^{n-1} \left(m_i^2 - E_{i,\textbf{k}}^2 \right) \left( \frac{E_{i,\textbf{k}} n_q}{\epsilon + P_0} - q_i \right) f^{(0)}_{i,\textbf{k}},
		\end{align}
	\end{widetext}
	and where we further impose Landau's definition of the frame \cite{landau1959course}, which leads to the additional constrain:
	\begin{align}
		&W^\mu = \sum\limits_{i \, = \, 1}^{N_\text{species}} \int_{\mathbb{R}^3} \mathrm{d}K_i \, E_{i,\textbf{k}} k^{\langle \mu \rangle}_i f^{(1)}_{i,\mathbf{k}} \overset{!}{=} 0 \quad \Rightarrow \nonumber\\
		&\quad  \sum\limits_{i \, = \, 1}^{N_{\text{species}}}  \sum\limits_{m \, = \, 0}^{M} \lambda^{(i)}_{m,q} \int_{\mathbb{R}^3} \mathrm{d} K_i \, E^{m+1}_{i,\textbf{k}} \left( m_i^2 - E_{i,\textbf{k}}^2 \right) f^{(0)}_{i,\textbf{k}} \overset{!}{=} 0 \label{eq:EnergyFluxLandau} .
	\end{align}
	Above we introduced the truncation order $M$; for the sake of simplicity the order is fixed to $M = 1$ which corresponds to the 14-moment approximation \cite{Denicol2012b}. We further define the corresponding conductivities as, $\sigma_{qq^\prime}/T = \kappa_{qq^\prime}/T^2$ and note that for $q = \mathrm{Q}$ or $q^\prime = \mathrm{Q}$ they are equivalent to the cross-electric conductivities introduced in Ref. \cite{Rose:2020sjv}. Especially, $\kappa_{QQ}/T^2 = \sigma_{\mathrm{el}}/T$ is the electric conductivity, which was already evaluated in various models \cite{Arnold:2000dr,Brandt:2012jc,Torres-Rincon:2012sda,Finazzo:2013efa,Amato:2013naa,RMarty:2013xph,Aarts:2014nba,Greif:2014oia,Puglisi:2014sha,PUGLISI:2015PLB,Brandt:2015aqk,Rougemont:2015ona,Ding:2016hua,Greif:2016skc,L.Thakur:17PRD,Rougemont:2017tlu,Greif2017PRL,Hammelmann:2018ath,Soloveva:2019xph,Fotakis:2019nbq,Rose:2020sjv}.
	
%----------------------------------------------------------------------	
	\subsection{Relaxation time approximation}
	\label{subsec:RTA}
	
	Anderson and Witting proposed an approximation to the collision term by defining a governing relaxation time \cite{ANDERSON1974466}. To first order, we write for each particle species $i$:
	\begin{align}
		\sum\limits_{j \, = \, 1}^{N_{\text{species}}} \mathcal{C}_{ij}^{(1)}[f_{i,\textbf{k}}] = -\frac{E_{i,\mathbf{k}}}{\tau_{i}} \left( f_{i,\mathbf{k}} - f^{(0)}_{i,\mathbf{k}}\right) = \nonumber \\ -\frac{E_{i,\mathbf{k}}}{\tau_{i}} f^{(1)}_{i,\mathbf{k}} + \mathcal{O}(\mathrm{Kn}^2) .
	\end{align}
	The relaxation time $\tau_i$ is related to the scattering rate $\Gamma_{i}(\mathbf{k}_i, T, \lbrace \mu_q \rbrace )$. For binary scattering we may write down the momentum dependent on-shell relaxation time \cite{Braaten:1991jj,Thoma:1993vs,Chakraborty:2010fr}:
	\begin{align}
		  &\frac{1}{\tau_{i} (\mathbf{k}_i,T, \lbrace \mu_q \rbrace )} = \Gamma_{i}(\mathbf{k}_i, T, \lbrace \mu_q \rbrace ) = \nonumber \\
		  &\sum\limits_{j \, = \, 1}^{N_{\text{species}}} \frac{1}{2!} \frac{1}{E_{i,\mathbf{k}}}\sum\limits_{a,b \, = \, 1}^{N_{\text{species}}}  \int_{\mathbb{R}^3} \mathrm{d} K^\prime_j \mathrm{d} P_a \mathrm{d} P^\prime_b  \, f^{(0)}_{j,\mathbf{k^\prime}} \tilde{f}^{(0)}_{a,\mathbf{p}} \tilde{f}^{(0)}_{b,\mathbf{p^\prime}} W^{ij \rightarrow ab}_{k k^\prime \rightarrow p p^\prime} . \label{eq:MomDepRelaxTimes}
	\end{align}
	From this we can also define the momentum-averaged relaxation time $\tau_{i,0}$ which may be used instead:
		\begin{align}
	&\frac{1}{ \tau_{i,0}(T, \lbrace \mu_q \rbrace ) }  = \Gamma_{i,0}(T, \lbrace \mu_q \rbrace ) \equiv \nonumber \\
	&\frac{1}{n_i} \int_{\mathbb{R}^3} \mathrm{d} K_i \, E_{i,\mathbf{k}} \Gamma_{i}(\mathbf{k}_i, T, \lbrace \mu_q \rbrace ) f^{(0)}_{i,\mathbf{k}}
	\end{align}
	with the on-shell particle density of species $i$:
	\begin{align}
		n_i(T,\lbrace \mu_q \rbrace) \equiv \int_{\mathbb{R}^3} \mathrm{d} K_i \, E_{i,\mathbf{k}} f^{(0)}_{i,\mathbf{k}} .
	\end{align}
This is also known as the relaxation time approximation (RTA) (cf. \cite{Hosoya:1983xm,Chakraborty:2010fr,Albright:2015fpa,Gavin:1985ph}). 

	In the classical limit and for the case of elastic, binary processes with constant isotropic cross sections, $\sigma_{\mathrm{tot}}^{ij \rightarrow ij} \equiv \sigma_{\mathrm{tot}} = \mathrm{const.}$, using Eq. \eqref{eq:ElasticTransitionRate} we can make the usual approximation (see e.g. \cite{Fotakis:2019nbq}):
	\begin{align}
		\frac{1}{\tau_{i,0} (T, \lbrace \mu_q \rbrace )} = \Gamma_{i,0}(T, \lbrace \mu_q \rbrace ) \approx \sigma_{\mathrm{tot}} \sum_j n_{j} = \sigma_{\mathrm{tot}} n_{\mathrm{tot}}.
	\end{align}
	Following Ref. \cite{Fotakis:2019nbq}, the diffusion coefficient matrix in the RTA can be expressed as:
	\begin{align}
		\kappa_{qq^\prime} = \frac{1}{3}  \sum\limits_{i \, = \, 1}^{N_{\text{species}}} \, q_i \int_{\mathbb{R}^3} \mathrm{d}K_i \, \tau_{i,0}(T,\mu_\mathrm{B}) \frac{1}{E_{i,\mathbf{k}}} \left( m_i^2 -  E^2_{i,\mathbf{k}} \right) \nonumber \\
		\times \left( \frac{E_{i,\mathbf{k}} n_{q^\prime}}{\epsilon + P_0} - q^\prime_i \right) f^{(0)}_{i,\mathbf{k}} \tilde{f}^{(0)}_{i,\mathbf{k}}. \label{eq:DiffMatrRTA}
	\end{align}

%----------------------------------------------------------------------	
	\subsection{Dynamical quasi-particle model for the quark-gluon plasma}
	\label{subsec:DQPM}
	
	In the dynamical quasi-particle model (DQPM) \cite{Peshier:2005pp,Cassing:2007nb,Cassing:2007yg,Cassing:2008nn,Soloveva:2019xph} the properties 
of the QGP are described in terms of strongly interacting dynamical quasi-particles -
quarks and gluons - with medium-adjusted properties. 
Their properties are constructed such that the equation of state (EoS) from lattice  Quantum  Chromo Dynamics (lQCD) is reproduced above  the  deconfinement  
temperature $T_c$.
	These quasi-particles are characterized by broad spectral functions $\rho_i$ ($i = q, {\bar q}, g$), which are assumed to have a Lorentzian form 
	\cite{Cassing:2007nb,Cassing:2007yg,Linnyk:2015rco}. They depend on the parton masses $m_i$ and their associated widths $\gamma_i$,
	\begin{align}
		\rho_i(\omega,{\bf p}) =
		\frac{\gamma_i}{\tilde{E}_{i,\mathbf{p}}} \left(
		\frac{1}{(\omega-\tilde{E}_{i,\mathbf{p}})^2+\gamma_i^2} - \frac{1}{(\omega+\tilde{E}_{i,\mathbf{p}})^2+\gamma_i^2}
		\right)\
		\label{eq:rho}.
	\end{align}
	Here, we introduced the off-shell energy $\tilde{E}_{i,\mathbf{p}} = \sqrt{ {\bf p}^2+M_i^2-\gamma_i^2 }$. In the DQPM the effective (squared) coupling constant $g^2$ is assumed to depend on temperature $T$ and baryon-chemical potential $\mu_\mathrm{B}$ \cite{Berrehrah:2013mua,Berrehrah:2014ysa,Berrehrah:2015ywa,Berrehrah:2016vzw}. At $\mu_\mathrm{B} = 0$ its temperature-dependence is parameterized via the entropy density $s(T,\mu_\mathrm{B} = 0)$ from lattice QCD from Refs. \cite{Borsanyi:2012cr,Borsanyi:2013bia} in the following way:
	\begin{equation}
	g^2(T,\mu_\mathrm{B} = 0) = d \Big( \left(s(T,0)/s^\mathrm{QCD}_{\mathrm{SB}} \right)^e -1 \Big)^f,
	\label{coupling_DQPM}
	\end{equation}
	with the Stefan-Boltzmann entropy density $s_{\mathrm{SB}}^{\mathrm{QCD}} = 19/9\ \pi^2 T^3$ and the dimensionless parameters $d = 169.934$, $e = -0.178434$ and $f = 1.14631$.
	In order to obtain the coupling constant at finite baryon chemical potential $\mu_\mathrm{B}$, we use of the 'scaling hypothesis' which assumes that $g^2$ is a function of the ratio of the effective temperature $T^* = \sqrt{T^2+ \mu^2_B/(9\pi^2)}$ and the $\mu_\mathrm{B}$-dependent critical temperature $T_c(\mu_\mathrm{B})$ as \cite{Berrehrah:16}:
	\begin{equation}
	g^2(T/T_c,\mu_\mathrm{B}) = g^2\left(\frac{T^*}{T_c(\mu_\mathrm{B})},\mu_\mathrm{B} =0 \right),
	\label{coupling}
	\end{equation}
	with $T_c(\mu_\mathrm{B}) = T_c(\mu_\mathrm{B} = 0) \sqrt{1-\alpha \mu_\mathrm{B}^2}$, $T_c(\mu_\mathrm{B} = 0) \approx 0.158$ GeV and $\alpha = 0.974\ \text{GeV}^{-2}$.
The $(T,\mu_\mathrm{B})$ behaviour of the DQPM coupling $g^2/(4 \pi)$ is shown in Fig. \ref{fig:g2tmuB}
in Appendix \ref{sec:ParticleProperties}. At $\mu_\mathrm{B}=0$ one can see a good agreement between the lQCD evaluation of the QCD running coupling $\alpha_s=g^2/(4 \pi)$ for $N_f=2$ \cite{Kaczmarek:2005PRD} and the DQPM running coupling.

	With the coupling $g^2$ fixed from lQCD, one can now specify the dynamical quasi-particle mass (for gluons and quarks) which is assumed to be given by the HTL thermal mass in the asymptotic high-momentum regime by \cite{Bellac:2011kqa,Linnyk:2015rco}
	\begin{align}
	& m^2_{g}(T,\mu_\mathrm{B})=\frac{g^2(T,\mu_\mathrm{B})}{6}\left(\left(N_{c}+\frac{N_{f}}{2}\right)T^2 
	+\frac{N_c}{2}\sum_{q}\frac{\mu^{2}_{q}}{\pi^2}\right), \nonumber \\
	 &  m^2_{q(\bar q)}(T,\mu_\mathrm{B})=\frac{N^{2}_{c}-1}{8N_{c}}g^2(T,\mu_\mathrm{B})\left(T^2+ \frac{\mu^{2}_{q}}{\pi^2}\right) \label{M9},
	\end{align}
	where $N_{c}=3$ the number of colors, while $N_{f} =3$ denotes the number of flavors. The strange quark has a larger bare mass which needs to be considered in its dynamical mass. Empirically we find $m_s(T,\mu_\mathrm{B})= m_{u/d}(T,\mu_\mathrm{B})+ \Delta m$ and $\Delta m$ =
	30 MeV.	Furthermore, the quasi-particles in the DQPM have finite widths, which are adopted in the form \cite{Berrehrah:16,PHSDreview}
	\begin{equation}
		\gamma_{i}(T,\mu_\mathrm{B}) = \frac{1}{3} C_i \frac{g^2(T,\mu_\mathrm{B})T}{8\pi}\ln\left(\frac{2c_m}{g^2(T,\mu_\mathrm{B})}+1\right),
	\end{equation}
where we use the QCD color factors for quarks, $C_q = C_F = \frac{N_c^2 - 1}{2 N_c} = 4/3$, and gluons, $C_g = C_A = N_c = 3$. Further, we fixed the parameter $c_m = 14.4$, which is related to the magnetic cut-off. We assume that the width of the strange quark is the same as that for the light ($u,d$) quarks. The evaluated masses and widths in the DQPM are shown in Fig. \ref{fig:DQPMmasses} in Appendix \ref{sec:ParticleProperties}.
	
	With the quasi-particle properties (or propagators) fixed as described above, one can
	evaluate the entropy density $s(T,\mu_\mathrm{B})$, the pressure $P_0(T,\mu_\mathrm{B})$ and energy
	density $\epsilon(T,\mu_\mathrm{B})$ in a straight forward manner by starting with
	the entropy density $s^{\mathrm{dqp}}$ and number density $n^{\mathrm{dqp}}$ in the propagator representation from Baym \cite{Baym,Blaizot:2000fc} and then identifying, $s = s^{\mathrm{dqp}}$ and $n_B = n^{\mathrm{dqp}}/3$ \cite{Soloveva:2019xph}. The isotropic pressure $P_0$ can then be obtained by using the Maxwell relation of a grand canonical ensemble:	
	\begin{align}
	P_0(T,\mu_\mathrm{B}) = & P_0(T_0,0)  + \int\limits_{T_{0}}^{T} s(T',0)\ dT' \nonumber \\
	& + \int\limits_{0}^{\mu_\mathrm{B}} n_B(T,\mu_\mathrm{B}')\ d\mu_\mathrm{B}' \label{pressure} ,
	\end{align}
	where the lower bound is chosen between $0.1 < T_0 < 0.15$ GeV.
	The energy density $\epsilon$ then follows from the Euler relation
	\begin{equation}
	\label{eps} \epsilon = T s - P +\mu_\mathrm{B} n_\mathrm{B} .
	\end{equation}
%	Further, the interaction measure is defined as:	
%	\begin{equation}
%	\label{wint} I \equiv \epsilon - 3P = Ts - 4P + \mu_\mathrm{B}\ n_\mathrm{B}
%	\end{equation}
%	which vanishes for massless and non-interacting degrees of freedom at $\mu_\mathrm{B} = 0$.
	
	In Ref. \cite{Pierre19} we found a good agreement between the entropy density $s(T)$, pressure $P_0(T)$,  energy density $\epsilon(T)$ and interaction measure $I(T) = \epsilon - 3P$ resulting from the DQPM, and results from lQCD obtained by the BMW
	group \cite{Borsanyi:2012cr,Borsanyi:2013bia} at $\mu_\mathrm{B} = 0$  and $\mu_\mathrm{B} = 400$ MeV.
	
	 From the above parametrizations of the masses, widths and the couplings, cross sections for anisotropic, inelastic binary tree-level QCD interactions with the dressed propagators and dressed couplings, have been evaluated which depend on temperature and baryon-chemical potential \cite{Moreau:2019vhw,Soloveva:2019xph}. 
	 The corresponding total cross sections are shown in Fig. \ref{dsigma_on_CM} in Appendix \ref{sec:ParticleProperties}, and are used of in the Chapman-Enskog evaluation described in Section \ref{subsec:CE}. Further, we provide new results for the complete diffusion coefficient matrix from the DQPM in the RTA by using Eq. \eqref{eq:DiffMatrRTA} and assuming relaxation times from Eq. \eqref{eq:MomDepRelaxTimes}. 
In the following the results from both approaches are presented. 
	 
%%%%%%%%%%%%%test case%%%%%%%%%%%%%%%%%%%%%%%%%%%%%%%%%%%%%%%%%%%%%%%%%%%%%%%%%%%%%%%

	\begin{figure*}
		\includegraphics[scale=1.0]{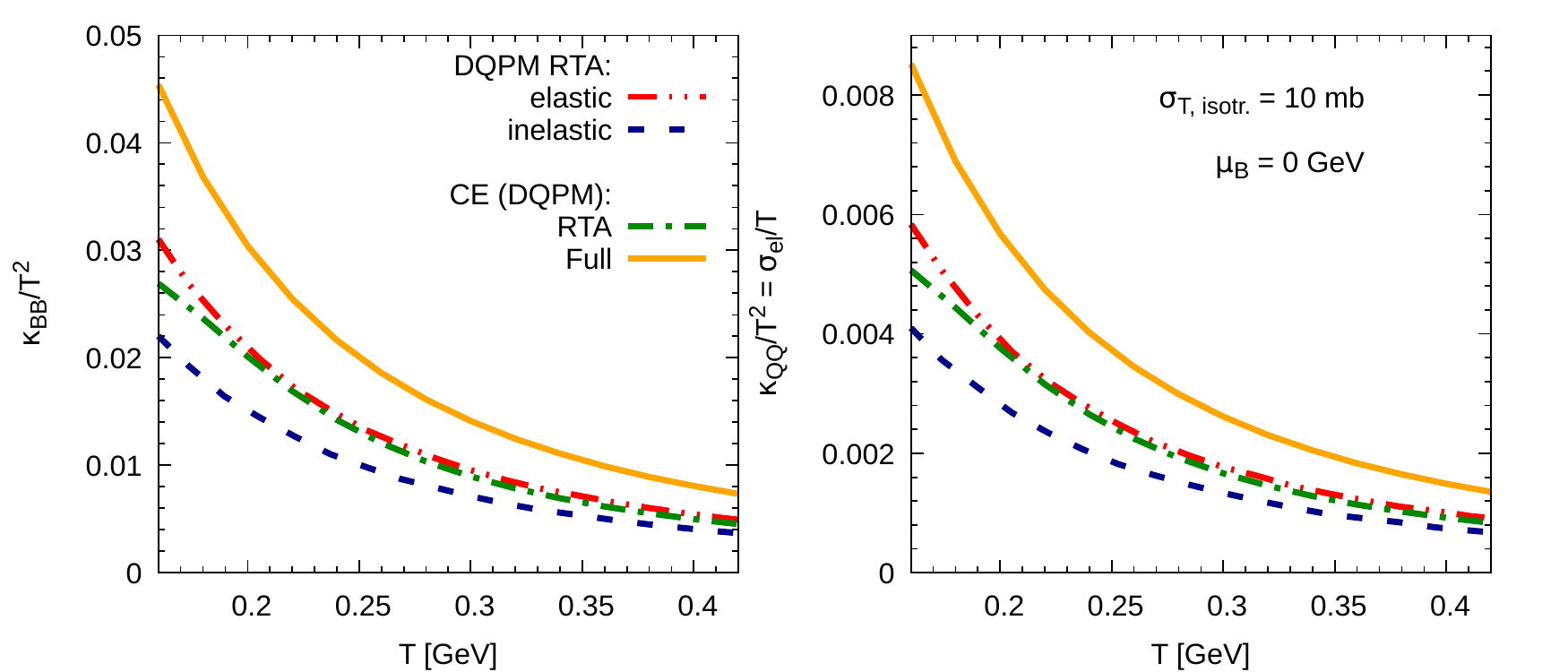}
		\caption{The scaled baryon diffusion coefficient, $\kappa_{\mathrm{BB}}/T^2$ (left), and the scaled electric conductivity, $\sigma_{\mathrm{el}}/T$ (right), for a partonic system with geometric cross sections, $\sigma_{\mathrm{tot}} = 10$ mb at vanishing baryon-chemical potential as a function of temperature from different approaches. We compare the DQPM RTA results from Eq. \eqref{eq:DiffMatrRTA} (for a system obeying quantum statistics, i.e. $a_i = \pm 1$ in Eq. \eqref{eq:Equilibrium}) with (blue dashed line) 
and without (red dashed-double-dotted line) inelastic, flavor-changing channels to the CE (DQPM) results either in RTA (green dashed-dotted line) from Eq. \eqref{eq:DiffMatrRTA} (for a system obeying classical statistics, i.e. $a_i = 0$ in Eq. \eqref{eq:Equilibrium}) or   for ``full" linearized collision term (orange solid line) from Eq. \eqref{eq:DiffMatrLin} ($a_i = 0$ in Eq. \eqref{eq:Equilibrium}).}
\label{fig:ConstCross}
    \end{figure*}

%--------------------------------------------------------------------------------	
	\section{Results}
	\label{sec:Results}
	
\begin{figure*}
		\centerline{ \includegraphics[width=75mm]{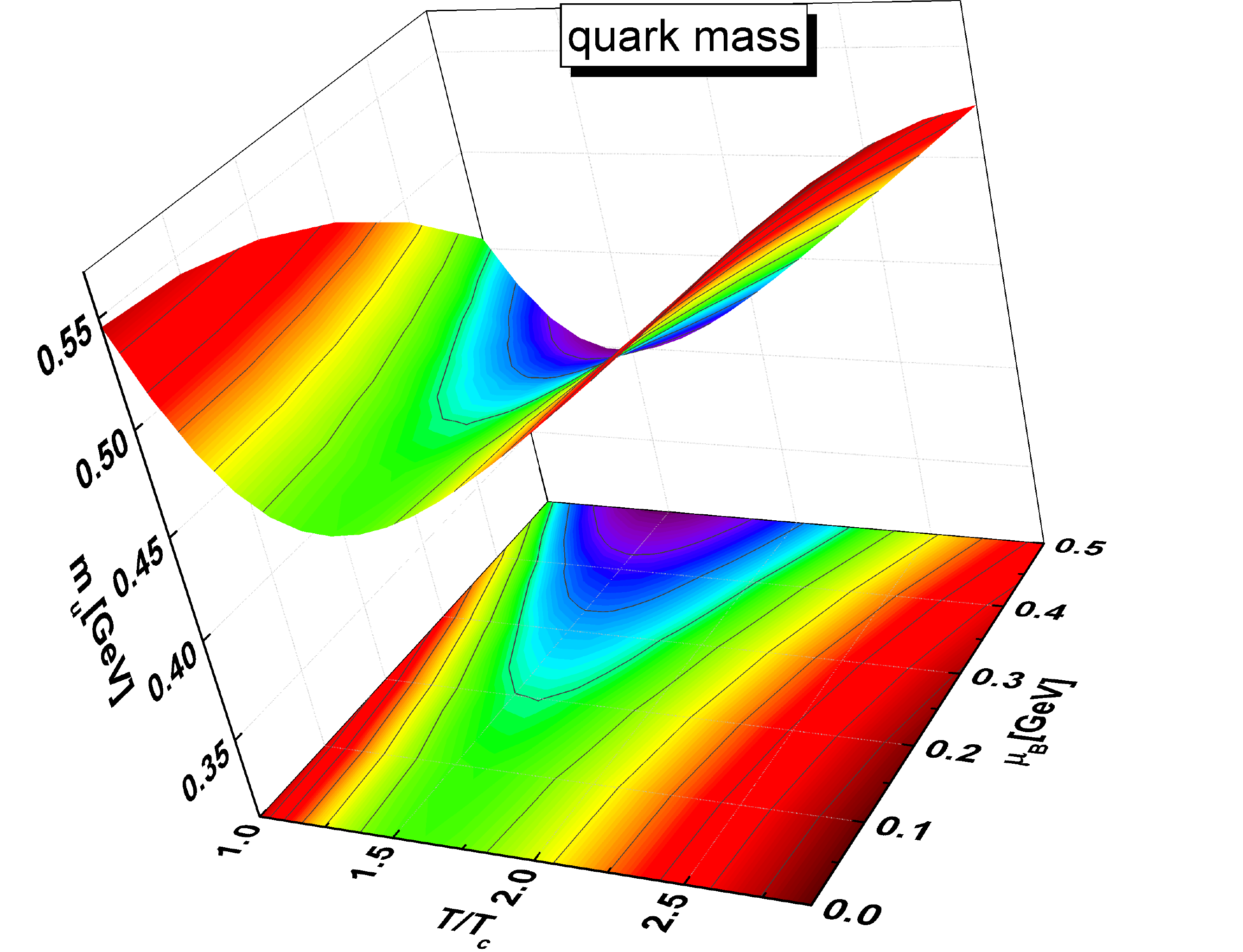}
			\includegraphics[width=75mm]{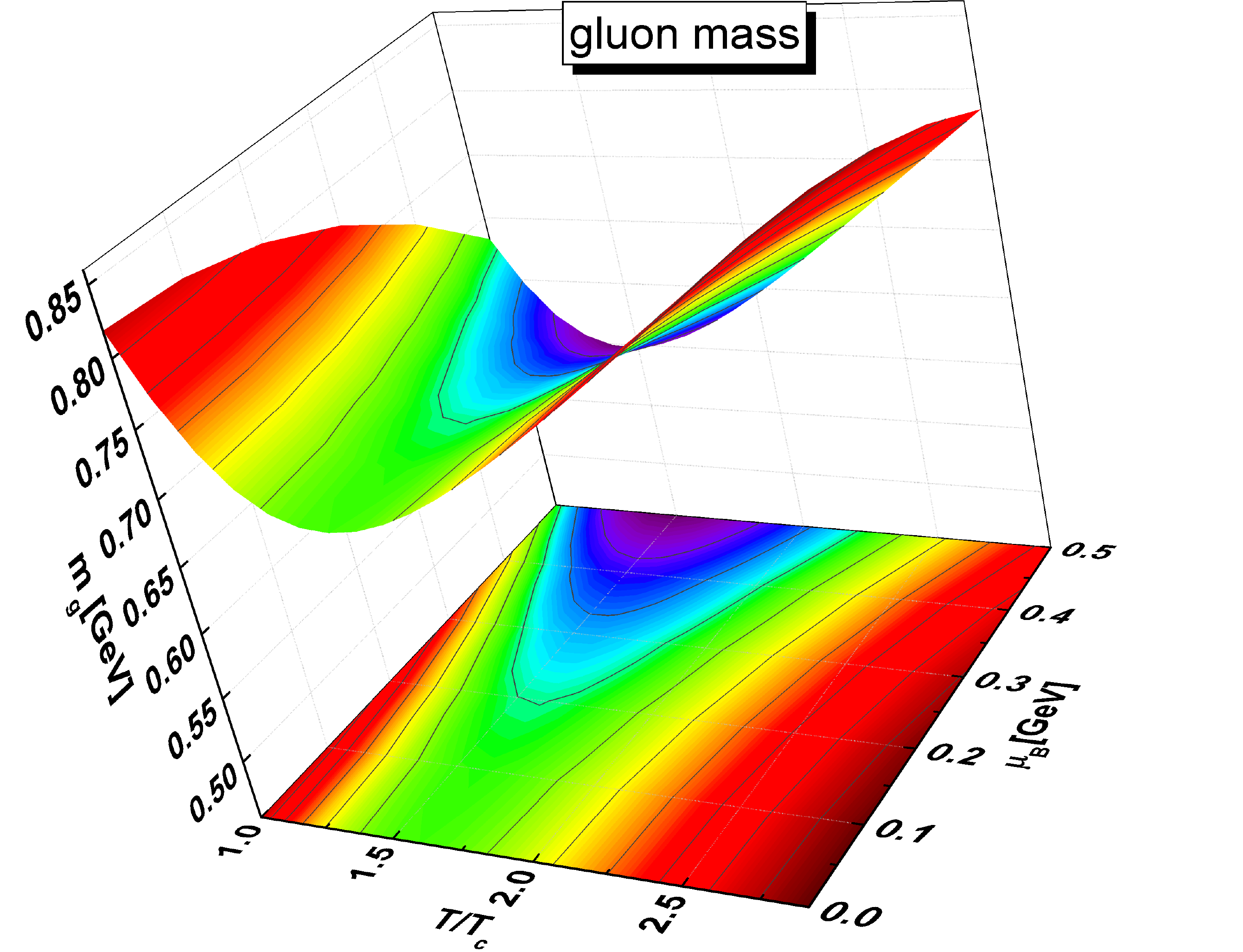}}
		\centerline{ \includegraphics[width=75mm]{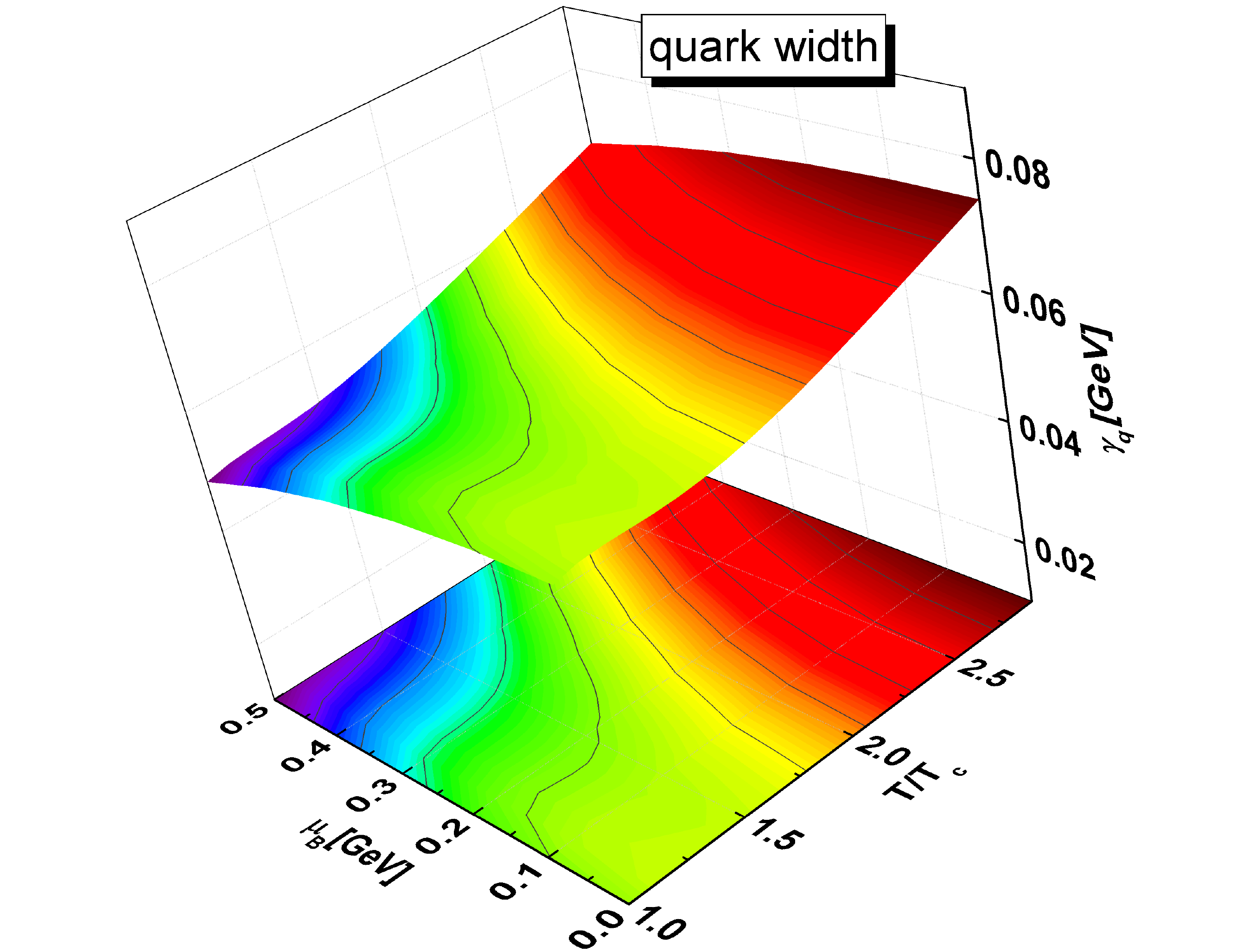}
			\includegraphics[width=75mm]{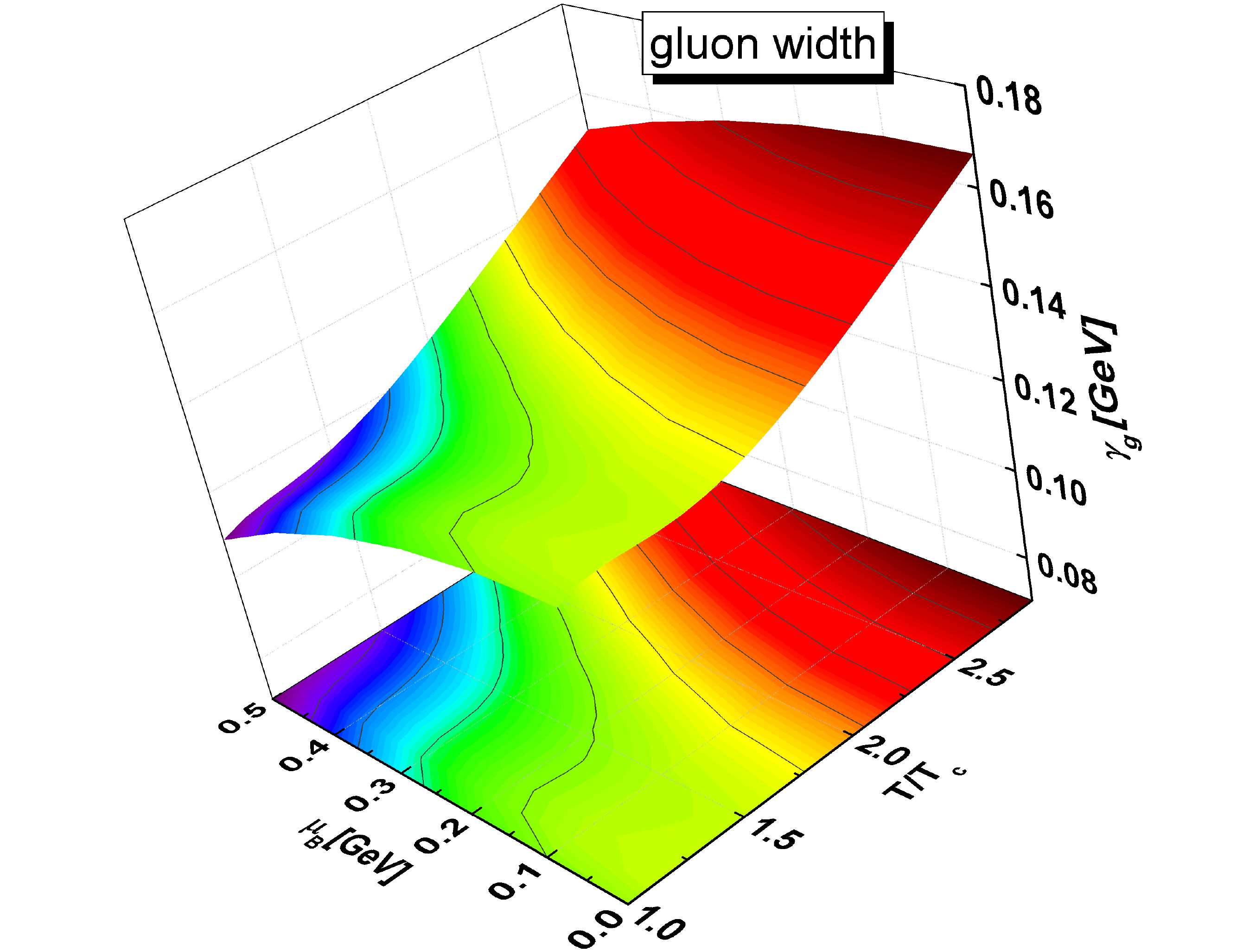}}
		%\vspace*{-0.15cm}
		
		\caption{The effective quark (\textit{left}) and gluon (\textit{right}) pole-masses $M$
			(\textit{upper row}) and their widths $\gamma$ (\textit{lower row}) from the actual DQPM 
			as a function of the temperature $T$ and baryon chemical potential $\mu_\mathrm{B}$ \cite{Soloveva:2020xof}. The strange quark mass was assumed to be $m_s = m_q + \Delta m$, with $\Delta m = 30$ MeV, and the width is identical with the widths of the light quarks, $\gamma_s = \gamma_q$. In the calculation of the diffusion matrix with the Chapman-Enskog method partons were assumed to be on-shell and thus the widths vanish.}
		\label{fig:DQPMmasses}
	\end{figure*}

	We provide first results for the diffusion coefficient matrix for the hot 
	quark-gluon plasma  at zero and finite baryon chemical potential $\mu_\mathrm{B}$ by applying the Chapman-Enskog method, reviewed in Sec. \ref{subsec:CE} and described in detail in Refs. \cite{Greif:2016skc,Greif2017PRL,Fotakis:2019nbq}, to 
a strongly interacting QGP system described by the DQPM (see Sec. \ref{subsec:DQPM} and Ref. \cite{Soloveva:2019xph}). This is meant to be a significant and important improvement to the 'simplified' model of a partonic system proposed in Refs. \cite{Greif2017PRL,Fotakis:2019nbq}. These results - obtained within the Chapman-Enskog method -  are further compared to the results for the diffusion coefficient matrix calculated within RTA approach based on the DQPM as well as  to various other models. The fact that the linearized Boltzmann equation is solved in the CE framework 
implies an improvement compared to approaches using the RTA (also see Ref. \cite{Fotakis:2019nbq}) 
in terms of accounting for high moments of the distribution function.
However, the proposed Chapman-Enskog method requires a few approximations  
for the QGP description, which are not in the spirit of the DQPM, in particular: 
	\begin{enumerate}
		\item The system is assumed to obey classical (Maxwell-Boltzmann) statistics (i.e. $a_i = 0$ for all particle species in Eq. \eqref{eq:Equilibrium}.		
		\item All particles are on-shell, therefore only the pole-masses from the DQPM, which depend on temperature and baryon-chemical potential, are assumed but their widths are neglected (for general quasiparticle properties see Appendix \ref{sec:ParticleProperties}, 
the Table \ref{tab:particleproperties} and the Fig. \ref{fig:DQPMmasses}).		
		\item Inelastic scattering channels are neglected. That implies that flavor-changing processes are not taken into consideration, i.e. $q\bar{q} \rightarrow q^\prime \bar{q}^\prime$ are not allowed.		
		\item All scattering processes are considered to be isotropic. We therefore feed total cross-sections into the CE evaluation which are evaluated from the anisotropic differential cross section from the DQPM via Eq. \eqref{sigma_on_CM}. The dependence on $\sqrt{s}$, temperature and baryon-chemical potential is taken into account, $\sigma^{ij \rightarrow ij}_{\mathrm{tot}} \equiv \sigma^{ij \rightarrow ij}_{\mathrm{tot}}(\sqrt{s},T,\mu_\mathrm{B})$ (see Appendix \ref{sec:ParticleProperties}, the Fig. \ref{fig:DQPMinteractions} for an example at $\mu_\mathrm{B} = 0$).
	\end{enumerate}
	We note that the CE method can in principle be improved such that approximations (1) and (3) become unnecessary. In Section \ref{subsec:ConstCross} we find indication that approximation (3) might have a non-neglectable impact. Such improvements are left for future work. 
However, the nature of the method makes further improvement of points (2) and (4)
difficult and require further detailed study.
	
	The explicit manifestations of the isotropic pressure, the energy density and the net charge densities in the CE evaluation are fed with the same lattice data to which the DQPM model was fitted to (see Section \ref{subsec:DQPM}) as far as they are available. Since the net strangeness and net electric densities are not available from the assumed lQCD results, we compute their values from kinetic theory (see Eq. \eqref{eq:chargedensities}).
	
In the following the results from the RTA approach applied to the original DQPM 
are denoted as ``DQPM RTA" while the results from the Chapman-Enskog method 
applied to the DQPM under assumptions (1)-(4) (as described above) is denoted 
as ``CE (DQPM)".
We remind here that for constant cross sections the scaled diffusion 
coefficients behave as $\kappa_{qq^\prime}/T^2 \sim 1/T^2$, 
as found in Ref. \cite{Fotakis:2019nbq}. 
Such a decreasing behavior is indeed found in Fig. \ref{fig:ConstCross}.
	    \begin{figure*}[t]
		\center{\includegraphics[scale=0.43]{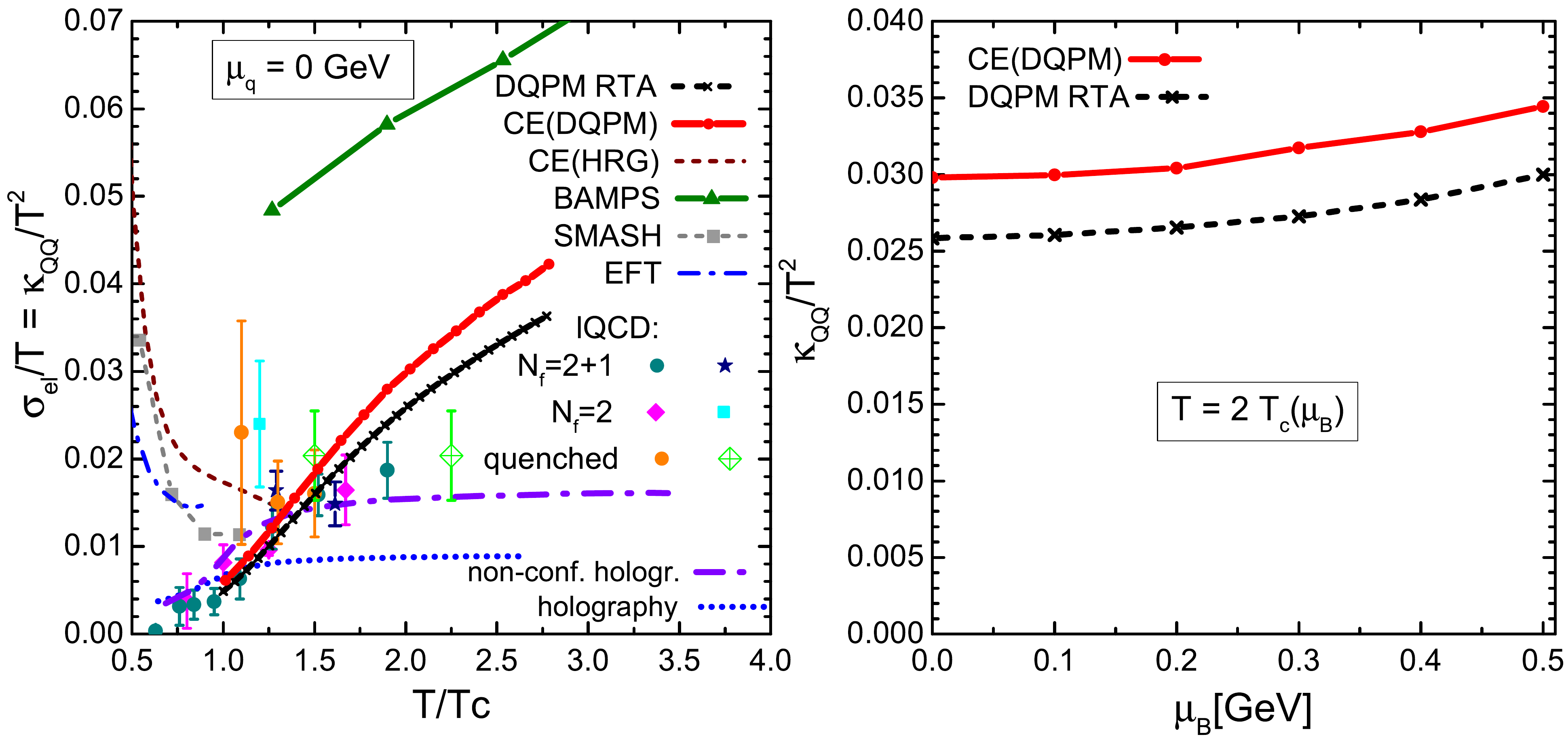}}
		\caption{ Left: Scaled electric conductivity, $\sigma_{\mathrm{el}}/T=\kappa_{\mathrm{QQ}}/T^2$, as a function of the scaled temperature $T/T_c$ at vanishing chemical potentials, $\mu_q = 0$, from various approaches. The results from the CE (DQPM) is shown by the red solid line, and from DQPM RTA -- by the black dashed line with crossed points. These are compared to results from the lattice QCD calculations (various shaped points with errorbars: quenched: orange circle-shaped points \cite{Ding:2016hua}, light green rhombus-shaped points \cite{Aarts:2007PRL}, $N_f=2:$ light cyan circle-shaped points \cite{Brandt:2012jc}, magenta rhombus-shaped points \cite{Brandt:2015aqk}, and $N_f=2+1:$ dark cyan circle-shaped points \cite{Aarts:2014nba} and blue stars \cite{Astrakhantsev:2019zkr}), the kinetic partonic cascade model BAMPS (dark-green solid line with triangular-shaped points) \cite{Greif:2014oia}, and from conformal \cite{Rougemont:2015ona} (blue dotted line) and non-conformal \cite{Finazzo:2013efa} (violet dashed-dotted line) holographic models. For temperatures below $T_c = 0.158$ GeV we show evaluations from hadronic models: SMASH \cite{Weil:2016zrk,Hammelmann:2018ath,Rose:2020sjv} (grey short-dashed line with squared points), effective field theory (EFT) \cite{Torres-Rincon:2012sda} (blue dashed-dotted line), and CE tuned to a hadron gas [CE (HRG)] from Refs. \cite{Greif:2016skc,Greif2017PRL,Fotakis:2019nbq} (dark-red dashed line).
Right: Scaled electric conductivity of the QGP at fixed scaled temperature, $T = 2 T_c (\mu_\mathrm{B})$, and vanishing $\mu_\mathrm{Q}$ and $\mu_\mathrm{S}$ are shown for varying baryon chemical potential $\mu_\mathrm{B}$ from the DQPM RTA (black dashed line with cross-shaped points) and the CE (DQPM) (red solid line with circle-shaped points) evaluation.}
		\label{fig:EC}
	\end{figure*}

%----------------------------------------------------------------------
	\subsection{Model study: constant isotropic cross sections}
	
	\label{subsec:ConstCross}
	
In order to evaluate the systematic differences between the DQPM RTA and 
the CE (DQPM) approaches we perform a here 'model study' by assuming  
 a (total) geometric cross section of $\sigma_{\mathrm{tot}} = 10$ mb for all interactions. For this comparison we consider the same assumptions as described in the preface above, but assume all channels to be characterized by the constant cross section. 
 
In Fig. \ref{fig:ConstCross} we show results for the scaled baryon coefficient, 
 $\kappa_{\mathrm{BB}}/T^2$ (left plot), and the scaled electric conductivity, 
 $\sigma_{\mathrm{el}}/T$ (right plot) at $\mu_\mathrm{B} = 0$ in a temperature range 
 from 160 MeV to 420 MeV.  The DQPM RTA calculations are presented for 
 two cases:  firstly, where all binary channels, including the inelastic ones, are considered (blue dashed line),  and for the case where only the elastic channels 
are accounted for (red dashed-double-dotted line). For the DQPM RTA results presented in this paper we use Eq. \eqref{eq:DiffMatrRTA} (for a system obeying quantum statistics, i.e. $a_i = \pm 1$ in Eq. \eqref{eq:Equilibrium}).
The CE (DQPM) calculations are presented in Fig.~\ref{fig:ConstCross} 
also for two cases:  in the first case we evaluate the coefficients in RTA with the help of Eq. \eqref{eq:DiffMatrRTA} (for a system obeying classical statistics, i.e. $a_i = 0$ in Eq. \eqref{eq:Equilibrium} ) 
under the assumption of the simplistic relaxation time, 
$\tau_0 = 1/{n_{\mathrm{tot}} \bar{\sigma}_T}$  (orange solid line), and for the second case we consider the full linearized Boltzmann equation via Eq. \eqref{eq:DiffMatrLin} (for a system obeying classical statistics, i.e. $a_i = 0$ in Eq. \eqref{eq:Equilibrium} ) (green dashed-dotted line).

	This 'model study' shows the influence of the consideration of the linearized Boltzmann equation compared to its relaxation time approximation, and the influence of the inelastic channels compared to its neglection. 
We find that the consideration of the full linearized collision term effectively reduces the scattering rate of a specific particle species, while in the RTA the scattering rate is overestimated. This is because in the collision term not only  the scattering of particles from a specific momentum bin into all other disjoint momentum bins is considered, but also the rescattering into this particular momentum bin is accounted for (gain and loss term). As argued in Ref. \cite{Fotakis:2019nbq} such an overestimation of the scattering rate leads to a decrease of the diffusion coefficients from RTA (which are anti-proportional to the rate).

Furthermore, we find that the inelastic channels lead to a further decrease of the  diffusion coefficients due to the repeated effective increase of the scattering rate as shown in Fig. \ref{fig:ConstCross}. Comparing the elastic version of the DQPM RTA evaluation with the CE (DQPM) calculation in the RTA limit (Eq. \eqref{eq:DiffMatrRTA}), we find a good agreement of the results at high temperatures. This is expected since the only difference between both calculations -- DQPM RTA
and  CE (DQPM) in the RTA limit -- is the consideration of quantum corrections and the more sophisticated (momentum-dependent) relaxation time in DQPM RTA. 

%----------------------------------------------------------------------
	\subsection{Diffusion coefficient matrix of the quark-gluon plasma}
	
	In the following we show results for the scaled diffusion coefficient matrix, $\kappa_{qq^\prime}/T^2$, for the partonic phase from the DQPM (RTA) and CE (DQPM) evaluation under the considerations described in the preface of Section \ref{sec:Results}. Additionally we consider two cases:
	\begin{itemize}
		\item We fix all chemical potentials to zero, $\mu_q = 0$ ($q=\mathrm{B}, \mathrm{S}, \mathrm{Q}$), 
		and show the temperature dependence of the coefficients.
		\item We fix the temperature to $T = 2 T_c(\mu_\mathrm{B})$, and show their dependence on the baryon chemical potential, $\mu_\mathrm{B}$. Here we further set the other chemical potentials to zero, $\mu_\mathrm{S} = 0$ and $\mu_\mathrm{Q} = 0$.
	\end{itemize}
	
	For most coefficients we find a rich $\mu_\mathrm{B}$-dependence. This dependence originates from the fact that all quarks carry baryon number and thus are sensitive to variations in $\mu_\mathrm{B}$. In Ref. \cite{Fotakis:2019nbq} the temperature dependence of these transport coefficients was reviewed, and it was found that they roughly 
scale as  $\kappa_{qq^\prime}/T^2 \sim 1/(\sigma_{\mathrm{tot}} T^2)$. 
In the case of the DQPM at fixed chemical potential, 
the cross sections depend on temperature as
$\sigma_{\mathrm{tot}} \sim 1/T^3$ or $\sim 1/T^4$ (for the considered temperature range). This depends on the combinations 
of $s -, t-, u-$ channels for different parton-parton scatterings: for $q-q$, $q-\bar{q}$ and $q-g$ scatterings $\sigma_{\mathrm{tot}} \sim 1/T^3$, while for the $g-g$ channel the terms $1/T^3$, $1/T^4$ have equivalent contribution to the total cross-section $\sigma_{\mathrm{tot}} \sim c_3/T^3+c_4/T^4$, where $c_3,c_4$ depend on $\sqrt{s}, \mu_\mathrm{B}$ (see e.g. Fig. \ref{fig:DQPMinteractions} in Appendix \ref{sec:ParticleProperties}). The temperature dependence of the cross-sections is in accordance with the temperature scaling of the DQPM coupling constant $g^2(T,\mu_\mathrm{B})$ (see e.g. Fig. \ref{fig:g2tmuB} in Appendix \ref{sec:ParticleProperties} ). This leads to a roughly quadratic dependence in temperature, $\kappa_{qq^\prime}/T^2 \sim T^2$, which is demonstrated in the figures below.
	
	We remind that the diffusion coefficient matrix is symmetric and therefore we may only show six instead of nine coefficients \cite{Onsager1931a,Onsager1931b}. In the following we subdivide the presentation of the conductivities in three sections: \textit{electric conductivities} ($\kappa_{\mathrm{QQ}}$, $\kappa_{\mathrm{QS}}$ and $\kappa_{\mathrm{QB}}$), \textit{strange conductivities} ($\kappa_{\mathrm{SS}}$ and $\kappa_{\mathrm{SB}}$), and finally the \textit{baryon conductivities} ($\kappa_{\mathrm{BB}}$). Since diffusion coefficients and conductivities are related to each other via temperature, $\kappa_{qq^\prime} = \sigma_{qq^\prime} T$, we use their denomination interchangeably.
	%%%%%%%%%%%%%%%%%%%%%%%%%%%%%%%%%%%%%%%%%%%%%%%%%%%%%%%%%%%%%%%%%%%%%%%%%%%%

	\begin{figure*}
		\includegraphics[scale=1.6]{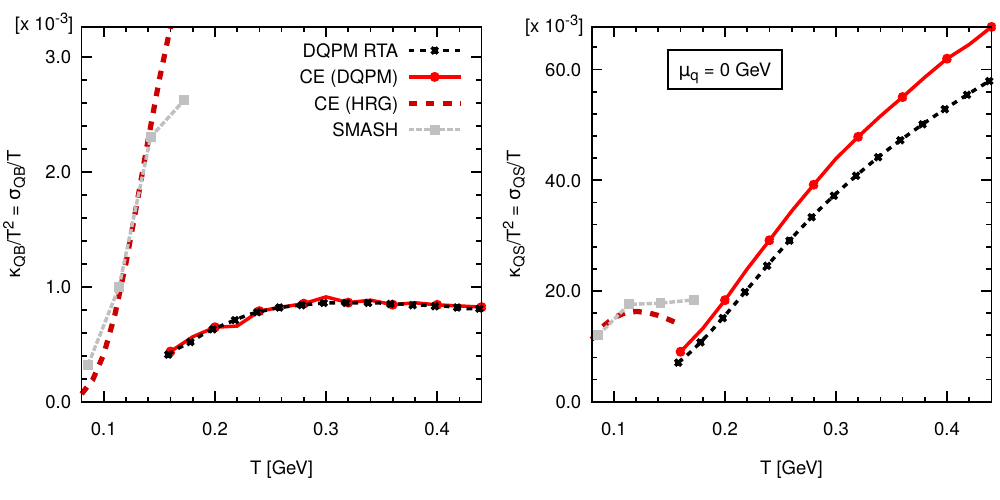}
		\caption{Scaled cross-electric conductivities, $\sigma_{\mathrm{QB}}/T^2$ (left) and $\sigma_{\mathrm{QS}}/T^2$ (right), from SMASH \cite{Weil:2016zrk,Rose:2020sjv} (grey short-dashed line with square-shaped points), the DQPM RTA (black dashed line with cross-shaped points), and the CE (DQPM) (red solid line with circle-shaped points) and CE (HRG) \cite{Greif2017PRL,Fotakis:2019nbq} (dark-red dashed line) evaluation at vanishing chemical potentials, $\mu_q = 0$, for temperatures between 80 and 420 MeV. \label{fig:multiECs}}
	\end{figure*}

	\begin{figure*}
		\includegraphics[scale=0.42]{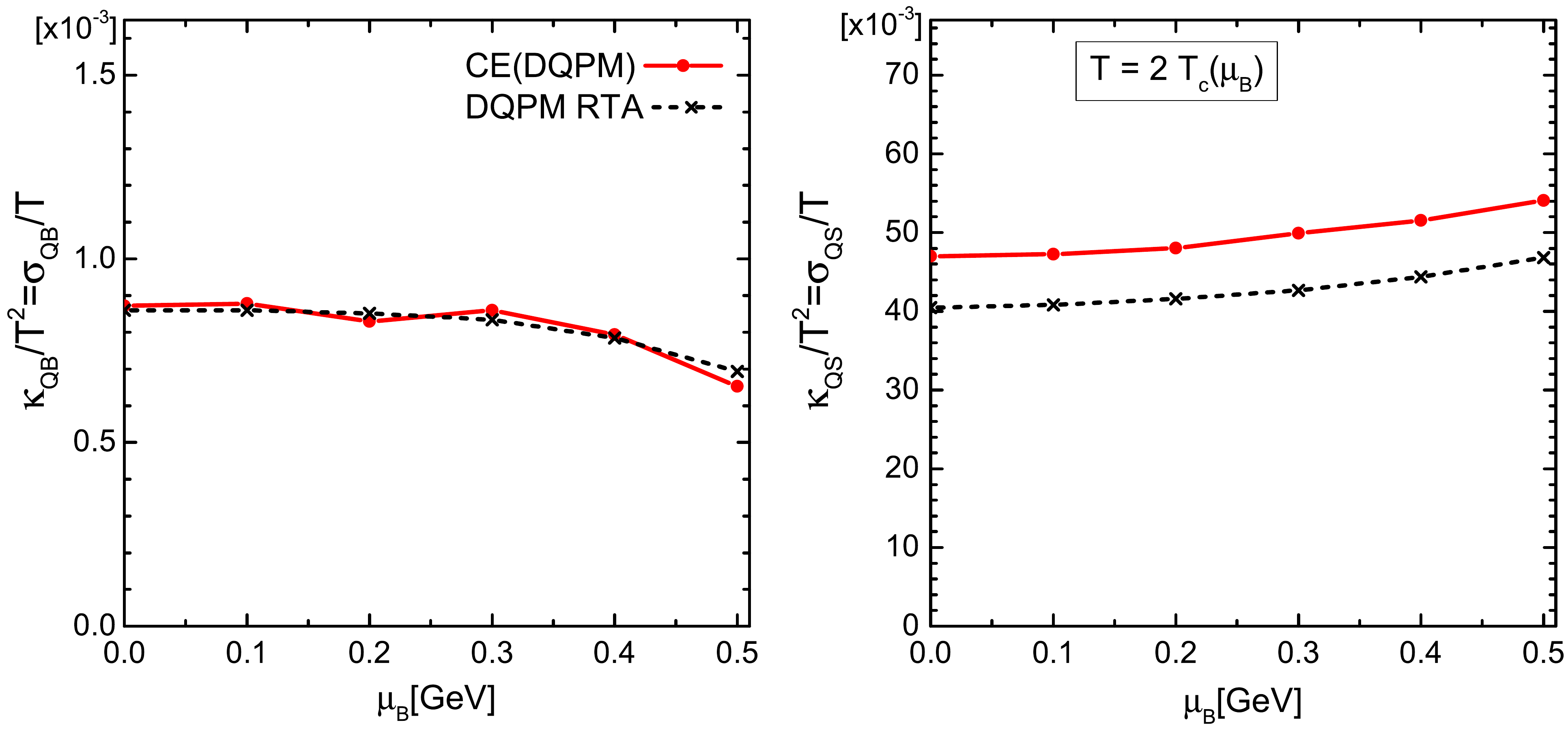}
		\caption{ Scaled cross-electric conductivities,  $\sigma_{\mathrm{QB}}/T^2$ (left) and $\sigma_{\mathrm{QS}}/T^2$ (right), from the DQPM RTA (black dashed line with cross-shaped points) and the CE (DQPM) evaluation at  fixed scaled temperature, $T = 2 T_c(\mu_\mathrm{B})$, shown over baryon chemical potential $\mu_\mathrm{B}$ in range 0 to 0.5 GeV. Further, the other chemical potentials are fixed to zero, $\mu_\mathrm{Q} = 0$ and $\mu_\mathrm{S} = 0$.  \label{fig:multiECs_muB}}
	\end{figure*}

%%%%%%%%%%%%%%%%%%%%%%%%%%%%%%%%%%%%%%%%%%%%%%%%%%%%%%%%%%%%%%%%%%%%%%%%%%%%	
%%%%%%%%%%%%%%%%%%%%strange%%%%%%%%%%%%%%%%%%%%%%%%%%%%%%%%%%%%%%%%%%%%%%%%%%%%%%%%

	\begin{figure*}
		\includegraphics[scale=0.42]{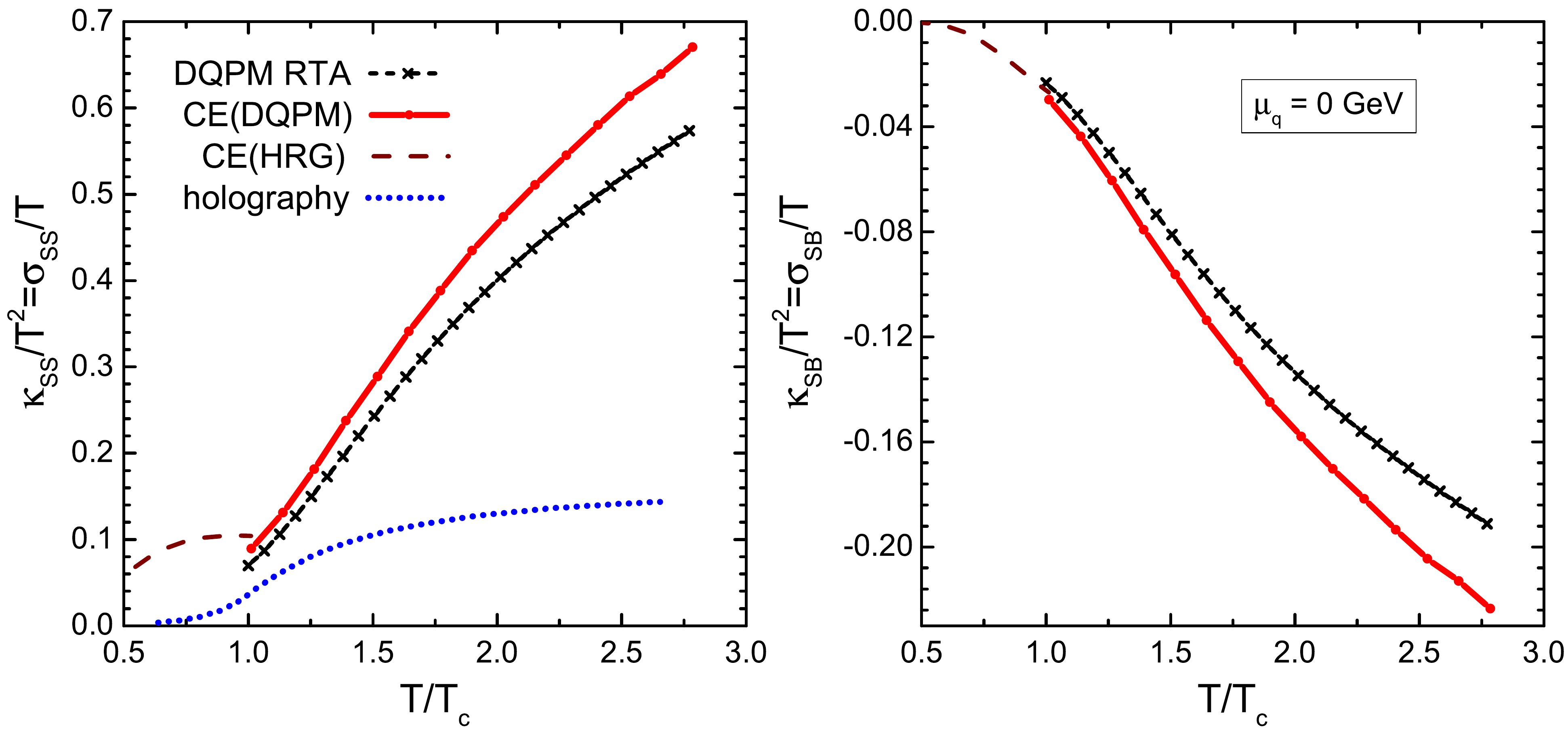}
		\caption{ Scaled strange and strange-baryon diffusion coefficients, $\kappa_{\mathrm{SS}}/T^2$ (left) and $\kappa_{\mathrm{SB}}/T^2$ (right), as a function of scaled temperature in the range from 0.5 to 3 $T_c$ at vanishing chemical potentials, $\mu_q = 0$. We compare results from CE (DQPM) (red solid line with circles), DQPM RTA (black dashed-line with crossed-shaped points), the CE (HRG) \cite{Greif2017PRL,Fotakis:2019nbq} (dark-red dashed line) and from conformal holography \cite{Rougemont:2015ona} (blue dotted line).  \label{fig:strange}}
	\end{figure*}

	\begin{figure*}
		\includegraphics[scale=0.42]{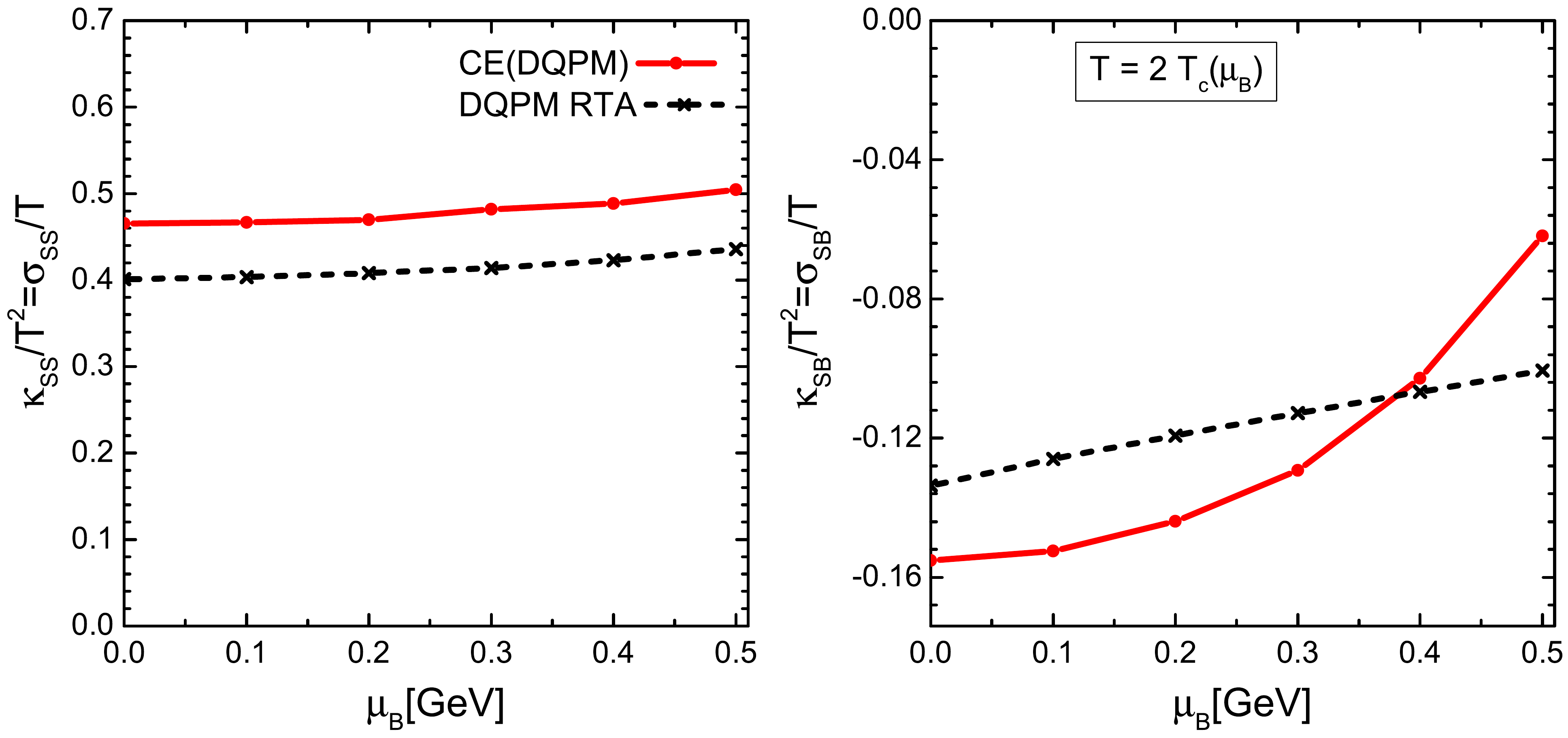}
		\caption{Scaled strange and strange-baryon diffusion coefficients, $\kappa_{\mathrm{SS}}/T^2$ (left) and $\kappa_{\mathrm{SB}}/T^2$ (right), from the DQPM RTA (black dashed line with cross-shaped points) and the CE (DQPM) evaluation at  fixed scaled temperature, $T = 2 T_c(\mu_\mathrm{B})$, shown over baryon chemical potential $\mu_\mathrm{B}$ in range 0 to 0.5 GeV. Further, the other chemical potentials are fixed to zero, $\mu_\mathrm{Q} = 0$ and $\mu_\mathrm{S} = 0$. \label{fig:strange_muB}}
	\end{figure*}
	
		%%%%%%%%%%%%%%%%%%%%-BB-%%%%%%%%%%%%%%%%%%%%%%%%%%%%%%%%%%%%%%%%%%%%%%%%%%%%%%%%

	\begin{figure*}
		\includegraphics[scale=0.42]{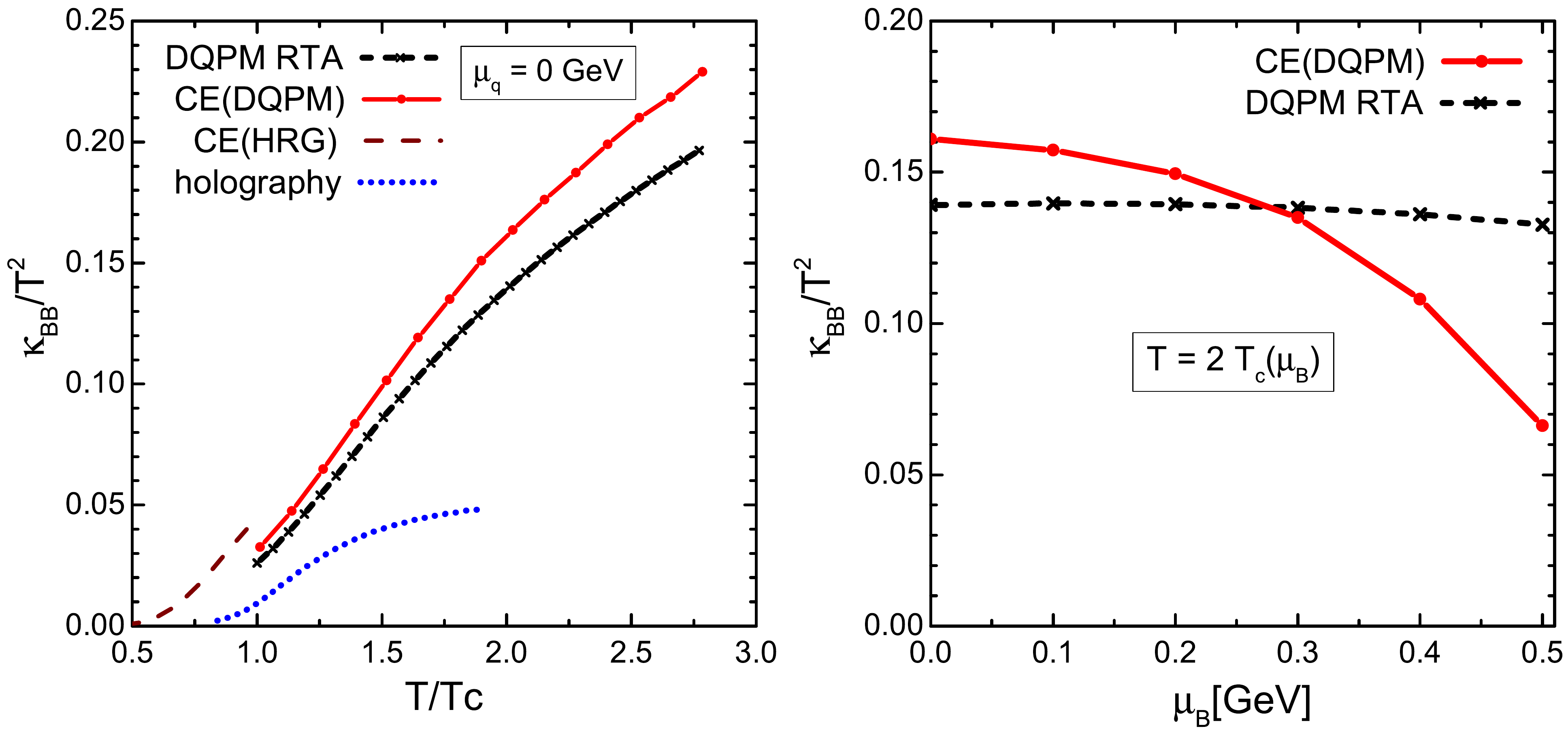}
		\caption{Scaled baryon diffusion coefficient, $\kappa_{\mathrm{BB}}/T^2$, (left) at vanishing chemical potentials, $\mu_q = 0$, as a function of the scaled temperature  from various approaches and (right) at  fixed scaled temperature, $T = 2 T_c(\mu_\mathrm{B})$, as a function of baryon chemical potential $\mu_\mathrm{B}$.  The strange and electric potential are fixed to zero: $\mu_\mathrm{S} = 0$ and $\mu_\mathrm{Q} = 0$. We show results from the CE evaluation tuned to DQPM (red solid line), as described above, and tuned to a hadron gas from Refs. \cite{Greif2017PRL,Fotakis:2019nbq} (dark-red dashed line). We again compare to the calculation from DQPM RTA \cite{Soloveva:2019xph} (black dashed line with crosses) and to conformal holography \cite{Rougemont:2015ona} (blue dotted line) as done for the electric conductivity.}
		\label{fig:BB}
	\end{figure*}

%%%%%%%%%%%%%%%%%%%%-BB-%%%%%%%%%%%%%%%%%%%%%%%%%%%%%%%%%%%%%%%%%%%%%%%%%%%%%%%%
%%%%%%%%%%%%%%%%%%%%%%%%%%%%%%%%%%%%%%%%%%%%%%%%%%%%%%%%%%%%%%%%%%%%%%%%%%%%
	\subsubsection{Electric conductivities}	
	
		The electric conductivity, $\sigma_{\mathrm{el}}/T$, was evaluated in various models (cf. Refs. \cite{Arnold:2000dr,Brandt:2012jc,Torres-Rincon:2012sda,Finazzo:2013efa,Amato:2013naa,RMarty:2013xph,Aarts:2014nba,Greif:2014oia,Puglisi:2014sha,PUGLISI:2015PLB,Brandt:2015aqk,Rougemont:2015ona,Ding:2016hua,Greif:2016skc,Berrehrah:16,L.Thakur:17PRD,Rougemont:2017tlu,Greif2017PRL,Hammelmann:2018ath,Soloveva:2019xph,Fotakis:2019nbq,Rose:2020sjv}). In Fig. \ref{fig:EC} we compare the results from DQPM RTA and CE (DQPM) to a variety of  models for both, the partonic \cite{Brandt:2012jc,Amato:2013naa,Aarts:2014nba,Greif:2014oia,Rougemont:2015ona,Finazzo:2013efa} and hadronic phase \cite{Hammelmann:2018ath,Rose:2020sjv,Torres-Rincon:2012sda,Greif:2016skc,Greif2017PRL,Fotakis:2019nbq}, at $\mu_\mathrm{q} = 0$ in a temperature range between 0 and 3$\,T_c$, where here the deconfinement temperature is $T_c = 158$ MeV. The Chapman-Enskog and RTA results for the dimensionless ratio of electric conductivity to temperature $\sigma_{\mathrm{el}}/T$ (later referred to as scaled electric conductivity) for $\mu_q = 0$ are presented in Fig.\ref{fig:EC} (left) as solid red and dashed black lines. Results for  both methods have a similar increase with temperature, which is mainly a consequence of the temperature dependence of the cross section (as discussed before) and also of the increasing total electric charge density \cite{Fotakis:2019nbq}.
	
	We find that the results from DQPM RTA and CE (DQPM) are consistent with results from lattice QCD in the vicinity of the crossover region, $1 \leq T/T_c \leq 1.5$. We again point out the apparent quadratic dependence on temperature which was shortly motivated in the preface of this section above. Due to our discussion from Section \ref{subsec:ConstCross} we suppose that a realistic result for the conductivities 
may be between the evaluations from DQPM RTA and CE (DQPM).

As follows from Fig. \ref{fig:EC}, the hadronic models presented there --
the hadronic transport model SMASH \cite{Weil:2016zrk,Hammelmann:2018ath,Rose:2020sjv} (grey short-dashed line with squared points), effective field theory (EFT) \cite{Torres-Rincon:2012sda} (blue dashed-dotted line), and CE tuned to a hadron gas [CE (HRG)] from Refs. \cite{Greif:2016skc,Greif2017PRL,Fotakis:2019nbq} (dark-red dashed line) -- substantially overestimate the lQCD data in the vicinity of $T_c$ as well as the results from  the conformal \cite{Rougemont:2015ona} (blue dotted line) and non-conformal \cite{Finazzo:2013efa} (violet dashed-dotted line) holographic models. The DQPM RTA results are in a good agreement in the vicinity of phase transition with the previous estimations for DQPM* from Ref. \cite{Berrehrah:16}, where non-relativistic formula for estimation the electric conductivity was used, which results in the linear dependence of the $\sigma_{el}/T$ on temperature while presented DQPM results show the quadratic dependence on temperature.

	Additionally to the electric conductivity, in Fig. \ref{fig:multiECs} we show the cross-electric conductivities, $\sigma_{\mathrm{BQ}}$ and $\sigma_{\mathrm{QS}}$, from the CE (DQPM) and the DQPM RTA calculation together with results achieved within 
SMASH \cite{Rose:2020sjv} and the CE (HRG) evaluation from Refs. \cite{Greif2017PRL,Fotakis:2019nbq} for the same thermal considerations for the hadronic phase. Comparing the results in both phases, we find a significant disagreement for $\sigma_{QB}$ around the crossover temperature. Further, we find such discrepancies to a smaller extend in the other electric conductivities and in the coefficients to follow. Such disagreement may hint to a difference in the chemical composition of the adjacent phases \cite{Rose:2020sjv}.

	Furthermore, as advertized in the preface, in Figs. \ref{fig:EC} (right) and \ref{fig:multiECs_muB} we present the sensitivity of the electric conductivities on $\mu_\mathrm{B}$ at fixed scaled temperature, $T = 2 T_c(\mu_\mathrm{B})$. Compared to the coefficients directly connected to the baryonic sector, we find a rather weak dependence on $\mu_\mathrm{B}$ (see discussion of $\kappa_{\mathrm{BB}}$ and $\kappa_{\mathrm{SB}}$). Surprisingly also $\sigma_{\mathrm{QB}}$ has such a weak dependence even though it also belongs to the baryonic sector. Further, $\sigma_{\mathrm{QB}}$ is very small - it has the smallest magnitude of all conductivities in the diffusion matrix. One can discuss its plausibility with a symmetry argument: 
	the $\sigma_{\mathrm{QB}}$ coefficient relates the generated electric current to the baryonic gradient which generates it (via the corresponding Navier-Stokes term). Assume a QGP with constant geometric cross section as discussed in Section \ref{subsec:ConstCross}. Further, assume that all quarks have the same mass. The down- and strange-quark have the same baryon number and electric charge, $\mathrm{B} = +1/3$ and $\mathrm{Q} = -1/3 e$, while the up-quark has $\mathrm{B} = +1/3$ and $\mathrm{Q} = +2/3e$, i.e. the same baryon number but an electric charge which is twice the magnitude but has the opposite sign (refer to Table \ref{tab:particleproperties}). Due to the quarks carrying the same baryon number, a baryon gradient generates a baryon current $V^\mu_\mathrm{B}$ which is equally composed by a current of up-, down- and strange-quarks ($V_i^\mu$): $V^\mu_\mathrm{B} = \sum_i \mathrm{B}_i V^\mu_i$, with $V^\mu_i = V^\mu_{\mathrm{quark}}$ $\forall i$. With this we can estimate the generated electric current:
	\begin{align}
		V^\mu_\mathrm{Q} = \sum_i \mathrm{Q}_i V^\mu_i = V^\mu_{\mathrm{quark}} \left( -\frac{1}{3} - \frac{1}{3} + \frac{2}{3} \right) = 0 .
	\end{align}
	The same argument can be made additionally accounting for the anti-quarks. The non-equal mass of the quarks and the varying cross sections lead to a non-vanishing $\sigma_{\mathrm{QB}}$. However, the above estimate illustrate the small magnitude of the respected coefficient.

	\subsubsection{Strange conductivities}
	
	We continue with the results for the strange sector: $\kappa_{\mathrm{SS}}$ and $\kappa_{\mathrm{SB}}$. The coefficient $\kappa_{\mathrm{SQ}}$, or equivalently $\sigma_{\mathrm{SQ}} = \sigma_{\mathrm{QS}}$, was already discussed above as part of the electric sector. Fig. \ref{fig:strange} shows the  $\kappa_{\mathrm{SS}}$ and $\kappa_{\mathrm{SB}}$ as a function of temperature at vanishing chemical potentials. Further, we show their $\mu_\mathrm{B}$-dependence in Fig. \ref{fig:strange_muB} in the range $\mu_\mathrm{B} = 0$ to 0.5 GeV at fixed scaled temperature, $T = 2 T_c(\mu_\mathrm{B})$ and for vanishing electric and strange chemical potential. We compare results from the DQPM RTA and CE (DQPM) computation to results from CE (HRG) in our recent publications \cite{Greif2017PRL,Fotakis:2019nbq}, and to results from conformal holography \cite{Rougemont:2015ona}.

We find that the baryon-strange diffusion coefficient is negative due to the definition of strangeness carried by the s-quark as has been already advocated in Ref. \cite{Greif2017PRL,Fotakis:2019nbq}. We obtain an almost quadratic dependence in temperature again, and a rather strong dependency on $\mu_\mathrm{B}$. However, the results from DQPM RTA for $\kappa_{\mathrm{SB}}$ in Fig. \ref{fig:strange_muB} show a slightly different $\mu_\mathrm{B}$-behavior than the results from CE (DQPM) for $\mu_\mathrm{B} \geq 0.3$ GeV. Furthermore, judging from Fig. \ref{fig:strange}, in the vicinity of the crossover region the results from CE (HRG) for the hadronic phase, and the calculation from the DQPM RTA and CE (DQPM) agree well.  As for other diffusion coefficients, the scaled strange diffusion coefficient from holography has a different temperature dependence and smaller values in the vicinity  of the crossover phase transition. 

%----------------------------------------------------------------------  	
	\subsubsection{Baryon conductivities}

In order to describe the deconfined QCD medium at the non-zero baryon density one should first consider the baryon diffusion coefficient $\kappa_{\mathrm{BB}}$.
	This diffusion coefficient was already evaluated in various models \cite{Arnold:2003zc,Rougemont:2015ona,Greif2017PRL,Ghiglieri:2018dib,Fotakis:2019nbq,Soloveva:2019xph}. Fig. \ref{fig:BB} (left) shows the temperature dependence of the baryon diffusion coefficient for the quark-gluon plasma estimated with the CE (DQPM) (red solid line) and DQPM RTA approaches (black dashed line with crosses). We also show the results from conformal holography \cite{Rougemont:2015ona}
(blue dotted lines). For temperatures below $T_c$ we again refer to the CE (HRG) calculation from Refs. \cite{Greif2017PRL,Fotakis:2019nbq} (dark-red dashed line). The comparison is presented at zero chemical potentials $\mu_q = 0$.
 Around $T_c$ the results from DQPM RTA, CE (DQPM) and CE (HRG) seem to be rather consistent with each other. Furthermore, we show its dependence on $\mu_\mathrm{B}$ at fixed scaled temperature $T = 2 T_c(\mu_\mathrm{B})$ in Fig. \ref{fig:BB} (right). %While the results from holography show almost no,  - NO holography on right plot!
The DQPM RTA shows a rather weak $\mu_\mathrm{B}$ dependence, while $\kappa_{\mathrm{BB}}$
from the CE (DQPM) decreases with $\mu_\mathrm{B}$.

	%%%%%%%%%%%%%%%%%%%%%%%%%%%%%%%%%%%%%%%%%%%%%%%%%%%%%%%%%%%%%%
	%%%%%%%%%%%%%%%%%%%%%%%%%%%%%%%%%%%%%%%%%%%%%%%%%%%%%%%%%%%%%%
	% C O N C L U S I O N
	%%%%%%%%%%%%%%%%%%%%%%%%%%%%%%%%%%%%%%%%%%%%%%%%%%%%%%%%%%%%%%
	%%%%%%%%%%%%%%%%%%%%%%%%%%%%%%%%%%%%%%%%%%%%%%%%%%%%%%%%%%%%%%
	\section{Conclusion}
	\label{sec:Conclusion}

In this study we have calculated the complete diffusion coefficient matrix  
$\kappa_{qq^\prime}$  ($q,q^\prime=\mathrm{B}, \mathrm{S}, \mathrm{Q}$) of the strongly interacting quark-gluon plasma 
by using the  Chapman-Enskog method as well as the  relaxation time approximation (RTA) from kinetic theory. We have explored the $T$ and $\mu_\mathrm{B}$ dependencies of the diffusion coefficients by considering  microscopical properties of quarks and gluons within the dynamical quasi-particle model (DQPM). The DQPM predictions of thermodynamic quantities for finite $\mu_\mathrm{B}$ show a good agreement with the available lQCD EoS \cite{Pierre19}. Moreover, for $\mu_\mathrm{B} =0$ the DQPM estimations of the QGP electric conductivity ($\sigma_{el}/T^2$) are in a good agreement with the $N_f=2+1$ lQCD results and in case of the specific shear and bulk viscosities ($\eta/s, \ \zeta/s$) the estimations are remarkably close to the predictions from the gluodynamic lQCD calculations \cite{Soloveva:2019xph}. \\
We find that the electric conductivities ($\kappa_{\mathrm{QQ}}$, $\kappa_{\mathrm{QS}}$ and $\kappa_{\mathrm{QB}}$), strange conductivities ($\kappa_{\mathrm{SS}}$ and $\kappa_{\mathrm{SB}}$), and finally the baryon conductivity ($\kappa_{\mathrm{BB}}$) have a similar temperature dependence in the vicinity of the phase transition while the $\mu_\mathrm{B}$ dependence is rather different among the considered diffusion coefficients. In particular, the diffusion coefficients $\kappa_{\mathrm{BB}}$ and $\kappa_{\mathrm{QB}}$ decrease with $\mu_\mathrm{B}$, while other coefficients increase. A suppression of baryon diffusion in a sQGP with finite $\mu_\mathrm{B}$  has been seen also in the holographic calculations \cite{Rougemont:2015ona}.

One of the main endeavors of this paper is to deliver reasonable estimates for the diffusion coefficients of the strongly interacting quark-gluon plasma. Furthermore, we compare the RTA evaluations from recent DQPM publications \cite{Soloveva:2019xph,Soloveva:2020xof,Soloveva:2020ozg} with  the Chapman-Enskog method. We demonstrate that once the cross sections and the (thermal) properties (masses, equation of state, etc.) of a system are known, the CE framework at hand is able to deliver consistent results. We show a good agreement for both methods with the available predictions from the literature for the partonic phase, in particular results for the scaled electric conductivity are remarkably close to the lQCD estimates at $\mu_\mathrm{B}=0$, as well as with the estimates for the hadronic phase. However, $\kappa_{QB}$ non-diagonal diffusion coefficient doesn't coincide well in the vicinity of the phase transition with the estimates for the hadronic phase, which can be interpreted as an indication of  a difference in the chemical composition of the adjacent phases. 
There are several model calculations of diagonal conductivities (mostly $\kappa_{\mathrm{QQ}}$) in the literature that are similar to
the RTA approach but used numerous restrictive assumptions for the evaluation of the relaxation times or cross-sections. Studying the diffusion coefficients of the QGP should have future  benefits  when  considering  hydrodynamical description for the time evolution of the deconfined QCD medium.

	%%%%%%%%%%%%%%%%%%%%%%%%%%%%%%%%%%%%%%%%%%%%%%%%%%%%%%%%%%%%%%
	%%%%%%%%%%%%%%%%%%%%%%%%%%%%%%%%%%%%%%%%%%%%%%%%%%%%%%%%%%%%%%
	% A P P E N D I X
	%%%%%%%%%%%%%%%%%%%%%%%%%%%%%%%%%%%%%%%%%%%%%%%%%%%%%%%%%%%%%%
	%%%%%%%%%%%%%%%%%%%%%%%%%%%%%%%%%%%%%%%%%%%%%%%%%%%%%%%%%%%%%%
	\newpage
	\section{Appendix}
	%%%%%%%%%%%%%%%%%%%%%%%%%%%%%%%%%%%%%%%%%%%%%%%%%%%%%%%%%%%%%%
	%%%%%%%%%%%%%%%%%%%%%%%%%%%%%%%%%%%%%%%%%%%%%%%%%%%%%%%%%%%%%%
	% P A R T I C L E   P R O P E R T I E S
	%%%%%%%%%%%%%%%%%%%%%%%%%%%%%%%%%%%%%%%%%%%%%%%%%%%%%%%%%%%%%%
	%%%%%%%%%%%%%%%%%%%%%%%%%%%%%%%%%%%%%%%%%%%%%%%%%%%%%%%%%%%%%%
	
	\subsection{Properties of partons in the DQPM}
	\label{sec:ParticleProperties}
	
	\begin{table}[h!]
		\begin{tabular}{|c|c|c|c|c|c|c|}
			\hline
			Name & Spin & Degeneracy & Baryon  & Electric & Strangeness \\
			     &      &            & Number  &   Charge &             \\
			\hline
			$g$	&	$1$		&	$16$	&	$0$		&	$0$	&	$0$ 	\\
			$u$	&	$1/2$		&	$6$	&	$+1/3$		&	$+2/3e$	&	$0$ 	\\
			$\bar{u}$	&	$1/2$		&	$6$	&	$-1/3$		&	$-2/3e$	&	$0$ 	\\
			$d$	&	$1/2$		&	$6$	&	$+1/3$		&	$-1/3e$	&	$0$ 	\\
			$\bar{d}$	&	$1/2$		&	$6$	&	$-1/3$		&	$+1/3e$	&	$0$ 	\\
			$s$	&	$1/2$		&	$6$	&	$+1/3$		&	$-1/3e$	&	$-1$ 	\\
			$\bar{s}$	&	$1/2$		&	$6$	&	$-1/3$		&	$+1/3e$	&	$+1$ 	\\
			\hline
		\end{tabular}
		\caption{Properties of the particle species considered in the  quark-gluon plasma in this work. Here, $e$ denotes the elementary electric charge in natural units. \label{tab:particleproperties}}
	\end{table}

%-------------

	\begin{figure*}
	\begin{minipage}[h]{0.45\linewidth}
		\center{\includegraphics[width=1\linewidth]{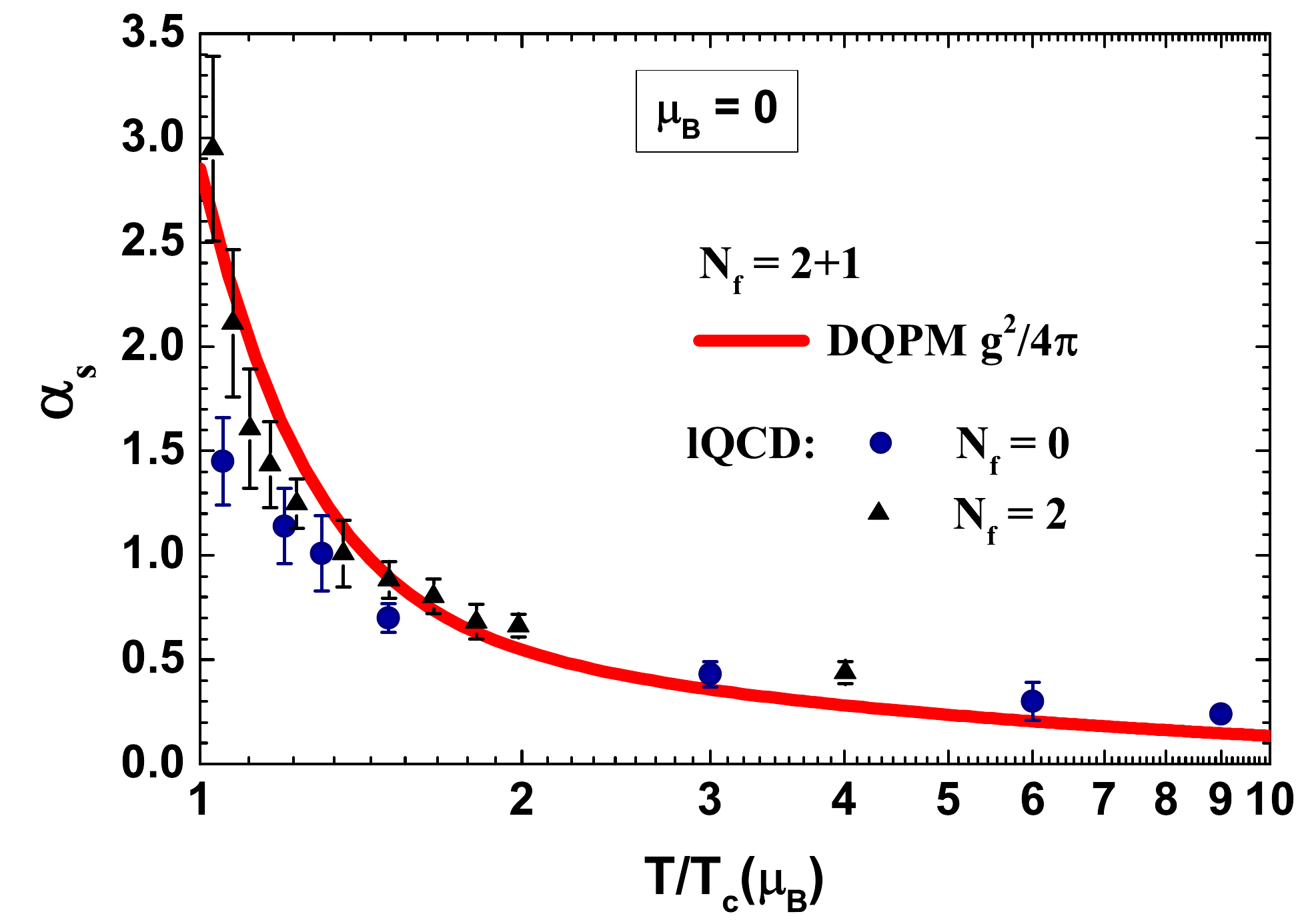} \\ a)}
	\end{minipage}
	\begin{minipage}[h]{0.45\linewidth}
		\center{\includegraphics[width=1\linewidth]{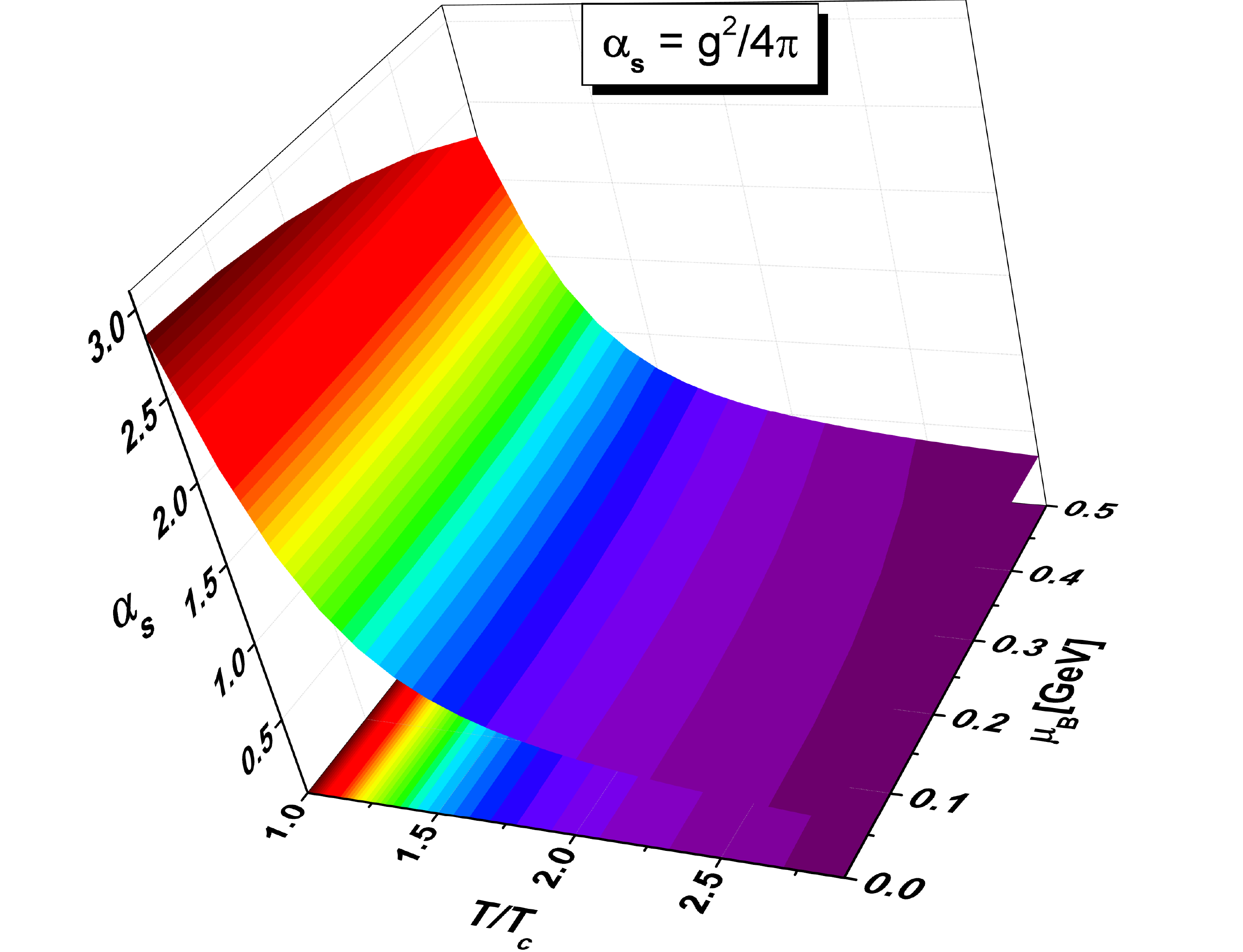}\\ b)}
	\end{minipage}
\caption{The running coupling $\alpha_s = g^2/(4 \pi)$ from the actual DQPM as a function of the scaled temperature $T/T_c$ at $\mu_\mathrm{B}=0$ a) and for moderate values of  baryon chemical potential $\mu_\mathrm{B}<= 0.5$ GeV (b) \cite{Soloveva:2020xof}. The lattice
results for quenched QCD, $N_f = 0$, (blue circles) are taken from Ref. \cite{Kaczmarek:2004gv} and for $N_f = 2$ (black triangles) are taken from Ref. \cite{Kaczmarek:2005PRD}.}
		\label{fig:g2tmuB}
	\end{figure*}

%----------------
	\begin{figure*}
	\begin{minipage}[h]{0.325\linewidth}
		\center{\includegraphics[width=1\linewidth]{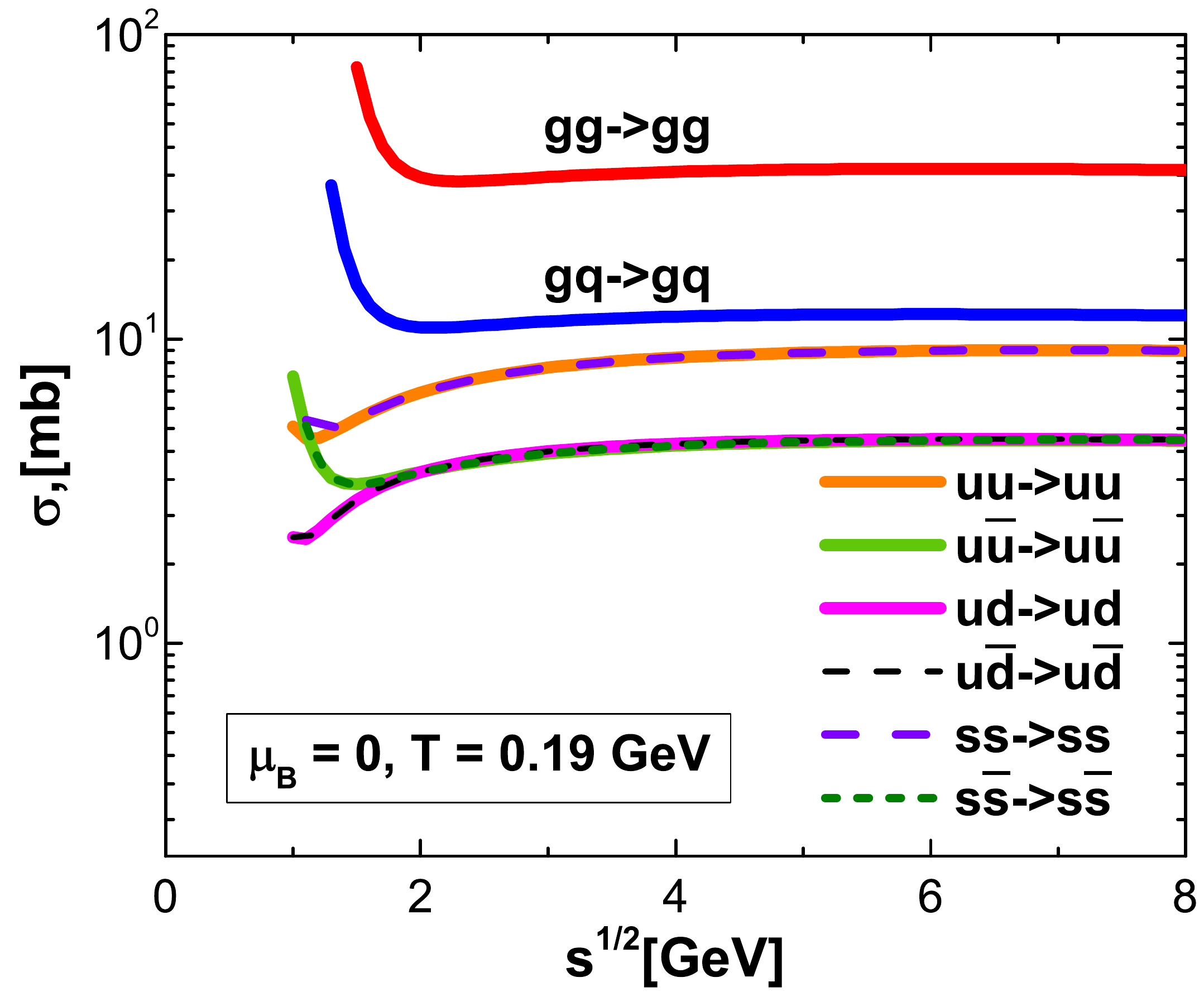} \\ a)}
	\end{minipage}
	\begin{minipage}[h]{0.325\linewidth}
		\center{\includegraphics[width=1\linewidth]{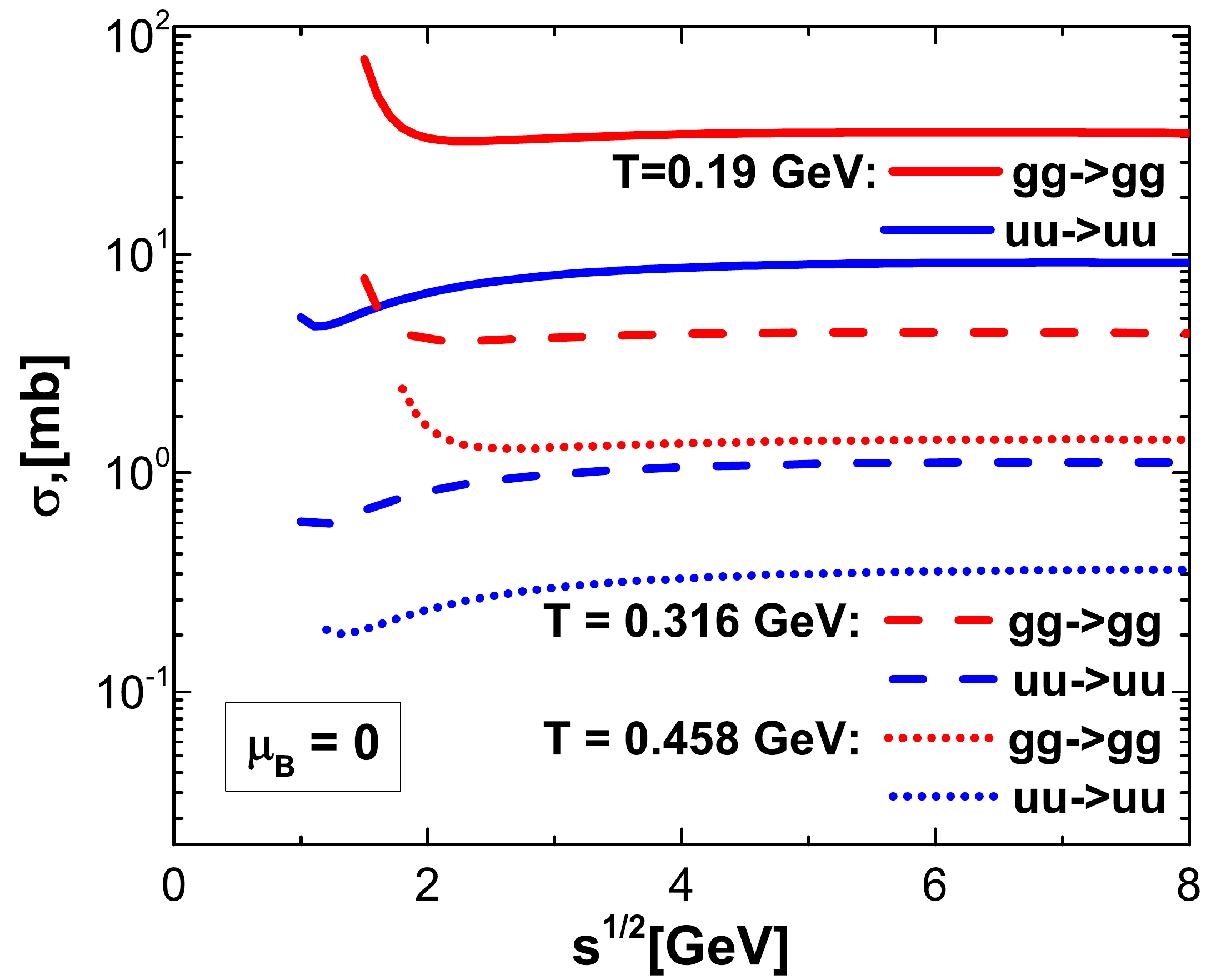}\\ b)}
	\end{minipage}
	\begin{minipage}[h]{0.325\linewidth}
		\center{\includegraphics[width=1\linewidth]{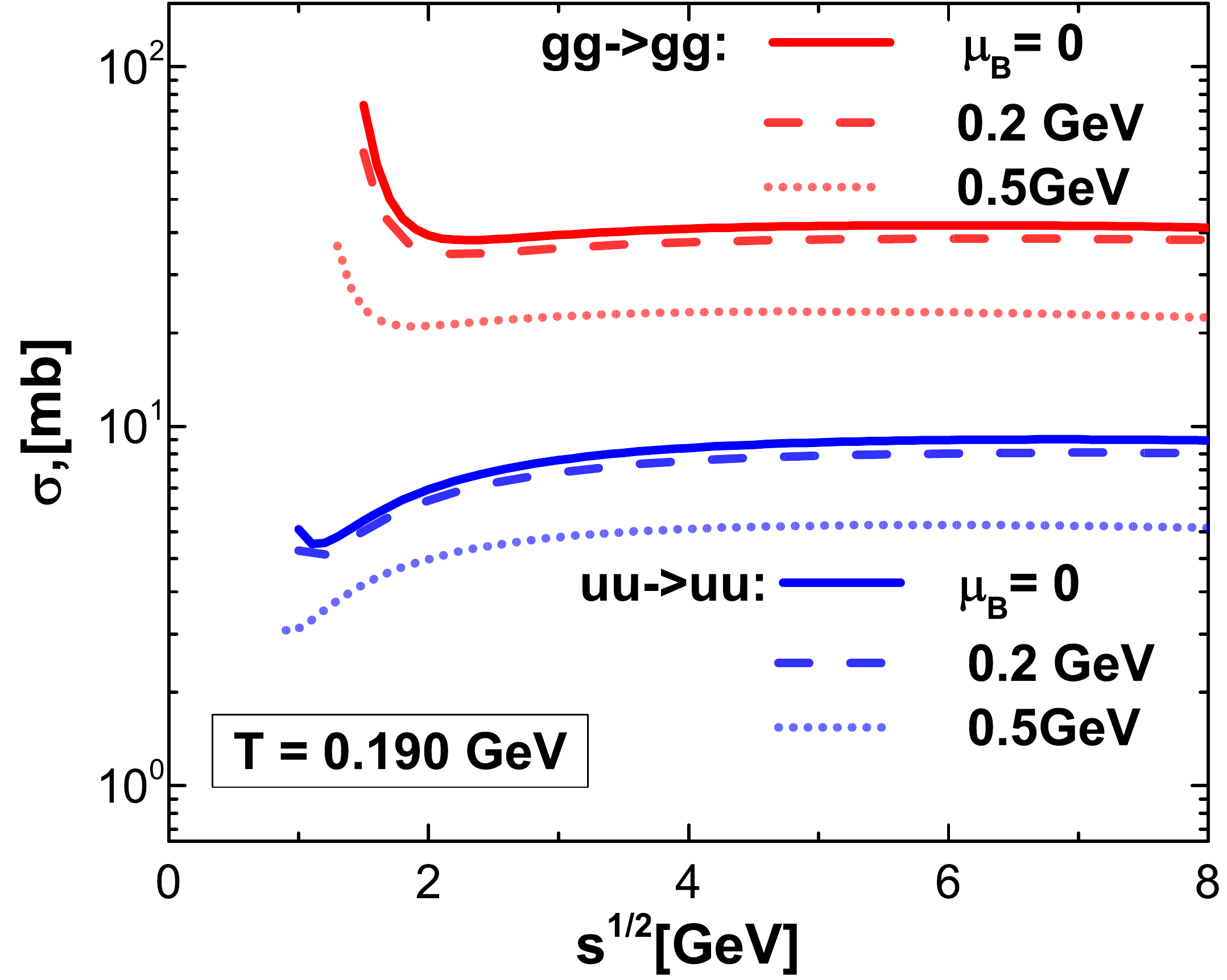} \\ c)}
	\end{minipage}
	\caption{DQPM total cross  sections between different partons for the on-shell case from Eq. \eqref{sigma_on_CM} evaluated in the center of mass of the collision system as a function of the collision energy $\sqrt{s}$ for a) $\mu_\mathrm{B}=0, T=0.19 $ GeV for the elastic channels, b) for $\mu_\mathrm{B}=0, T=0.19,0.316,0.458 $ GeV for both elastic and inelastic channels, and c) $\mu_\mathrm{B}=0,0.2,0.5 $ GeV for two examples. 
For the minimal allowed center-of-momentum energy of the colliding partons only their pole-masses are taken into account. In the Chapman-Enskog evaluation of the diffusion matrix, the inelastic channels were neglected for simplicity.}
	\label{fig:DQPMinteractions}
	\end{figure*}

	\newpage
	\section*{Acknowledgments}
The authors acknowledge inspiring discussions with H. van Hees, T. Song and J. M. Torres-Rincon. Also the authors acknowledge support by the Deutsche Forschungsgemeinschaft (DFG, German Research Foundation)
through the CRC-TR 211 'Strong-interaction matter under extreme conditions'– project number 315477589 – TRR 211. O.S. and J.A.F.	acknowledge support from the  Helmholtz Graduate School
for Heavy Ion research.
Furthermore, we acknowledge support by the Deutsche Forschungsgemeinschaft 
by the European Union’s Horizon 2020 research and innovation program under grant agreement No 824093 (STRONG-2020) and by the COST Action THOR, CA15213. 
The computational resources have been provided by
the LOEWE-Center for Scientific Computing and the "Green Cube" at GSI, Darmstadt.

	%%%%%%%%%%%%%%%%%%%%%%%%%%%%%%%%%%%%%%%%%%%%%%%%%%%%%%%%%%%%%
	%%%%%%%%%%%%%%%%%%%%%%%%%%%%%%%%%%%%%%%%%%%%%%%%%%%%%%%%%%%%%
	% B I B L I O G R A P H Y
	%%%%%%%%%%%%%%%%%%%%%%%%%%%%%%%%%%%%%%%%%%%%%%%%%%%%%%%%%%%%%
	%%%%%%%%%%%%%%%%%%%%%%%%%%%%%%%%%%%%%%%%%%%%%%%%%%%%%%%%%%%%%
	%\bibliographystyle{ieeetr} 
	\bibliographystyle{apsrev4-1}
	\bibliography{paper.bib} 
	
\end{document}